\documentclass[pra, amsfonts, amssymb, amsmath,doi,reprint, showkeys, twoside,nobibnotes,floatfix, nofootinbib]{revtex4-2}

\usepackage{epsfig}
\usepackage{amssymb}
\usepackage{amsthm}
\usepackage{tabularx}
\usepackage{braket}

\begin{document}

\title{Fourier-transform spectroscopy and relativistic electronic structure calculation \\ on the $c^3\Sigma^+$ state of KCs}

\author{Artis Kruzins$^{1}$}
\author{Valts Krumins$^{1}$}
\author{Maris Tamanis$^{1}$}
\author{Ruvin Ferber$^{1}$}
\author{\\Alexander V. Oleynichenko$^{2,3}$}
\author{Andr\'ei Zaitsevskii$^{2,3}$}
\author{Elena A. Pazyuk$^{3}$}
\author{Andrey V. Stolyarov$^{3}$}
\affiliation{$^{1}$Laser Center, Faculty of Physics, Mathematics and Optometry, University of Latvia, 19 Rainis blvd, Riga LV-1586, Latvia}
\affiliation{$^{2}$Petersburg Nuclear Physics Institute named by B. P. Konstantinov of National Research Center ``Kurchatov Institute'', 188300, Leningradskaya oblast, Gatchina, mkr. Orlova roscha, 1, Russia}
\affiliation{$^{3}$Department of Chemistry, Lomonosov Moscow State University, 119991, GSP-2, Moscow, Leninskie gory 1/3, Russia}
\email{ferber@latnet.lv}

\date{\today}

\begin{abstract}
The Ti:Saphire laser operated within 13800 - 11800 cm$^{-1}$ range was used to excite the $c^3\Sigma^+$ state of KCs molecule directly from the ground $X^1\Sigma^+$ state. The laser-induced fluorescence (LIF) spectra of the $c^3\Sigma^+ \rightarrow a^3\Sigma^+$ transition were recorded with Fourier-transform spectrometer within 8000 to 10000 cm$^{-1}$ range. Overall 673 rovibronic term values belonging to both $e/f$-components of the $c^3\Sigma^+(\Omega=1^{\pm})$ state of $^{39}$KCs, covering vibrational levels from $v$ = 0 to about 45, and rotational levels $J\in [11,149]$  were determined with the accuracy of about 0.01 cm$^{-1}$; among them 7 values  for $^{41}$KCs. The experimental term values with $v\in [0,22]$ were involved in a direct point-wise potential reconstruction for the $c^3\Sigma^+(\Omega=1^{\pm})$ state, which takes into account the $\Omega$-doubling effect caused by the spin-rotational interaction with the nearby $c^3\Sigma^+(\Omega=0^-)$ state. The analysis and interpretation were facilitated by the fully-relativistic coupled cluster calculation of the potential energy curves for the $B^1\Pi$, $c^3\Sigma^+$, and $b^3\Pi$ states, as well as of spin-forbidden $c-X$ and spin-allowed $c-a$ transition dipole moments; radiative lifetimes and vibronic branching ratios were calculated. A comparison of relative intensity distributions measured in vibrational $c-a$ LIF progressions with their theoretical counterparts unambiguously confirms the vibrational assignment suggested in [\emph{J. Szczepkovski, et. al.}, JQSRT, \textbf{204}, 133-137 (2018)].
\end{abstract}


\maketitle              

\section{Introduction}\label{intro}

\begin{figure}
\includegraphics[width=\columnwidth]{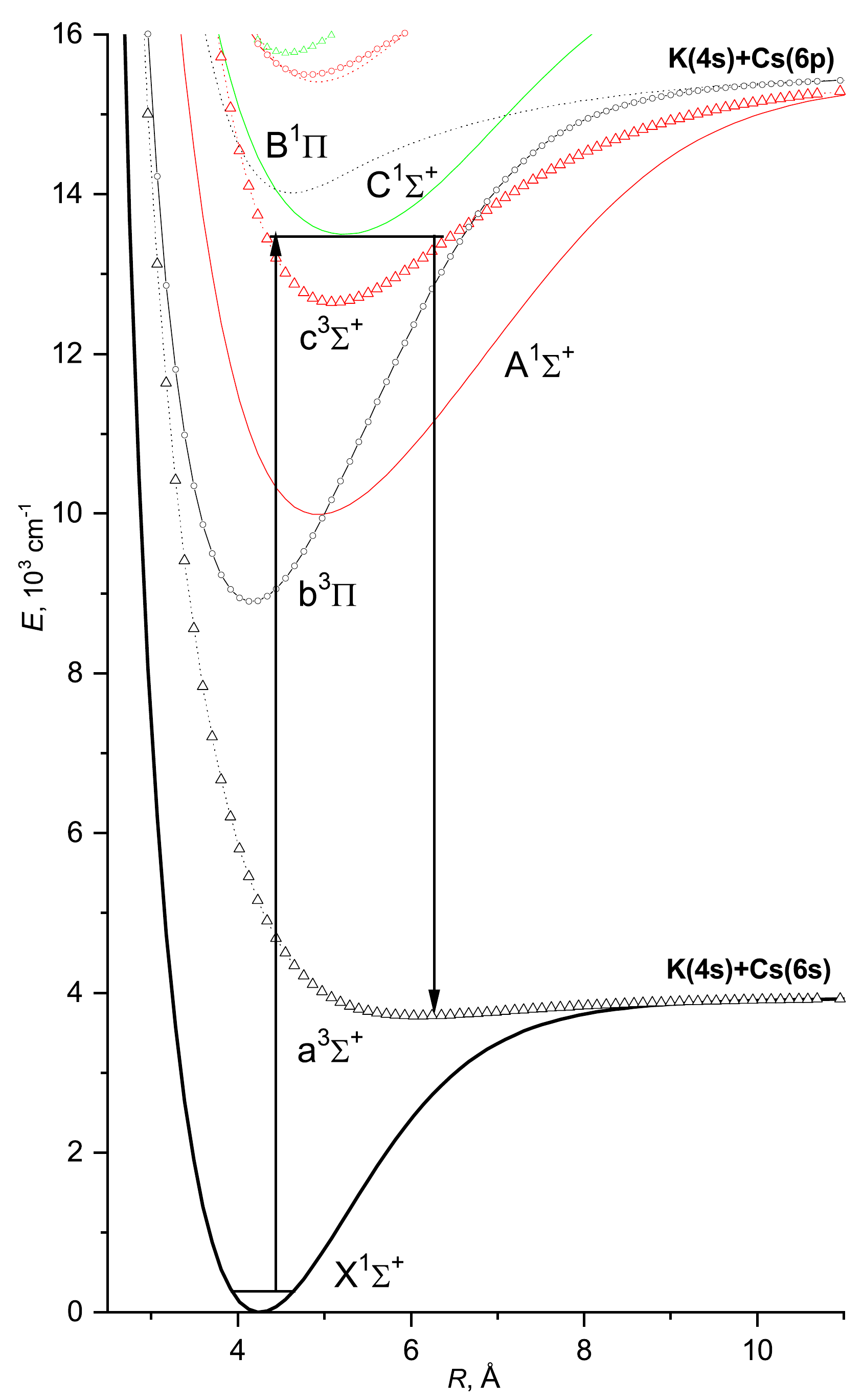}
\caption{Scheme of the Hund's coupling case ($a$)PECs for selected low-lying states of KCs taken from Ref.~\cite{Korek2000}. Vertical arrows illustrate optical excitation-observation scheme used in the present experiment.} \label{Fig1c3S}
\end{figure}

The low-lying excited electronic states of polar alkali diatomic molecules, especially the ones correlating to the first excited atomic asymptote, namely the $A^1\Sigma^+$, $b^3\Pi$, $c^3\Sigma^+$, and $B^1\Pi$ states (in Hund's coupling case $(a)$ notation), see Fig.~\ref{Fig1c3S}, are attracting a particular attention for various reasons. Due to singlet-triplet interaction, they are often used as intermediate states in transforming the cold and ultracold molecules to their low-lying ground state, preferably to the "absolute" rovibronic ground state $X^1\Sigma^+(v_{X}=0,J_{X}=0)$, suggesting a variety of applications, see~\cite{Carr, Quemener} for a review. Transfer of ultracold species to the deeply bonded ground state levels via first excited electronic states was realized for RbCs ~\cite{Kerman,Shimisaki}, LiCs~\cite{Deigtlmayr}, NaCs~\cite{Zabawa}, and LiRb~\cite{Stevenson}. Besides, the mutually perturbed states allow to access the higher-lying triplet states manifold by applying multi-step laser excitation technique, which is especially favorable for heteronuclear diatomics due to absence of the $u/g$ selection rule. At the same time, the states in question represent a quite challenging object to a spectroscopic study, especially for alkali diatomics containing a heavier atom Rb or Cs, first of all due to the pronounced spin-orbit (SO) and spin-rotational coupling effects~\cite{Bernath2005book} leading to numerous intramolecular perturbations both of local and global nature~\cite{Field2004book}.

The ultimate strategy would be to involve these strongly perturbed states into a rigorous coupled-channel deperturbation treatment~\cite{Pazyuk:19}, which would yield deperturbed interatomic potentials and corresponding non-adiabatic coupling functions allowing reproducing the experimental term values and transition strengths with a spectroscopic level of confidence~\cite{Pazyuk:15}. For Rb or Cs containing polar alkali diatomics a coupled-channels (CC) deperturbation analysis has been successfully applied to the $A^1\Sigma^+$ and $b^3\Pi$ states fully mixed by a strong SO interaction. The 4x4 coupled-channels model yielded deperturbed potentials for both $A$ and $b$ states, as well as the on-diagonal and off-diagonal SO functions, which allowed one to reproduce thousands of rovibronic term values of the isolated singlet-triplet $A-b$ complex with standard deviation of about 0.01 cm$^{-1}$, or even better, see~\cite{Docenko2007} for NaRb,~\cite{Zaharova2009} for NaCs, ~\cite{PRA2010,PRAdirect2010,PRA2013} for KCs,~\cite{Docenko2010,Kruzins2014} for RbCs, and~\cite{Alps2016} for KRb.

The first necessary step to involve two other nearby states, namely $c^3\Sigma^+$ and $B^1\Pi$, is to obtain accurate experimental term value data on these states in a wide enough energy range covering an abundant set of vibrational and rotational levels. In earlier works~\cite{Wang1992,Matsabura1993} the local SO interaction between $c^3\Sigma^+$ and $B^1\Pi$ states was studied in NaRb applying high-resolution sub-Doppler laser spectroscopy; the deperturbed molecular constants of each state and the relevant SO matrix elements were determined in the framework of the effective Hamiltonian approach. Polarization-labeling spectroscopy, as well as photoassociation (PA) and pulsed laser-depletion spectroscopy of ultracold species were applied to NaCs  in~\cite{Grochola2011} yielding an abundant data set of the $c^3\Sigma^+(\Omega = 1)$ state used to extract the conventional Dunham molecular constants, including the $v^{\prime},J^{\prime}$-independent $\Omega$-doubling constant $q\approx +1.4\times 10^{-4}$~cm$^{-1}$. The $B^1\Pi$ state of NaCs has been investigated in~\cite{Zaharova2007} in a wide energy range revealing numerous perturbation regions by high-resolution Fourier-transform spectroscopy (FTS) of laser-induce fluorescence (LIF) and in~\cite{Grochola2010} by polarization labeling spectroscopy; the empirical point-wise adiabatic potentials have been reconstructed applying the robust Inverted Perturbation Approach (IPA)~\cite{Pashov2000}. A reduced version of coupled vibrational channel was suggested and applied to $c^3\Sigma^+$ state of KRb in~\cite{Pazyuk:18}.

Performing spectroscopic studies on ultracold RbCs the authors of~\cite{Shimisaki,Bergeman2004} have suggested to consider the mutually perturbed $b^3\Pi$, $c^3\Sigma^+$, and $B^1\Pi$ states [2(1), 3(1), and 4(1) states (in Hund's coupling case ($c$) notation) as a $B-b-c$ complex of the strongly mixed singlet-triplet states assuming that the SO interaction with the nearby $A^1\Sigma^+$ state can be neglected. In these studies, the authors are separately considering the SO interaction inside the $b^3\Pi - c^3\Sigma^+$ and $c^3\Sigma^+ - B^1\Pi$ pairs. A preliminary analysis of local perturbations in the $c^3\Sigma^+$ state of RbCs caused by high levels of the $b^3\Pi$ state and low levels of the $B^1\Pi$ state has been performed in Ref.~\cite{Bergeman2004} basing on fragmentary data from $c \leftarrow a$ re-excitation spectra after a decay to the $a^3\Sigma^+$ state of the PA resonances of laser-cooled atoms. It should be noted that the low vibrational levels of the regularly perturbed $c^3\Sigma^+$ state clearly demonstrated the nonzero splitting $2\lambda\approx -9$~cm$^{-1}$ between the lower $c(\Omega = 1)$ and upper $c(\Omega = 0^-)$ components of the non-rotating molecule (so-called inverted $\rho$-doubling effect in the $^3\Sigma^+$ state~\cite{Watson1971}). The authors of ~\cite{Shimisaki} have realized the short-range PA via the strong spin-allowed $a \rightarrow c$ transition. The excited 'intermediate' PA state decays to the $X^1\Sigma^+$ state via the spin-forbidden triplet-singlet $c\to X$ transition, which allowed to efficiently produce the ultracold RbCs molecules in their rovibronic ground state $X^1\Sigma^+(v_X=0,J_X=0)$. The work was motivated by earlier investigation of the close-lying singlet $B^1\Pi$ state of RbCs~\cite{Birzniece2013}.

The present study is focused on both experimental and theoretical investigations of the $c^3\Sigma^+$ state of the closest analogue of RbCs, namely the KCs molecule. This molecule is currently under intensive research aimed to produce a stable ensemble of ultracold KCs species. The interspecies Feshbach resonances in collisions of ultracold K and Cs have been observed in~\cite{Grobner2017}. A number of optical schemes based on stimulated Raman adiabatic passage (STIRAP) were proposed in~\cite{Borsalino2016} to create ultracold KCs in its "absolute" ground state, starting from a weakly bound level of the ground state manifold. In this connection, a detailed experimental and theoretical information on the lowest excited electronic states of KCs is very helpful. As far as spectroscopic studies of the mixed $B - b - c$ states are concerned, the $B^1\Pi$ state of KCs has been studied in~\cite{Birzniece2012,Birzniece2015}, where 1174 term values of the $B^1\Pi$ state were obtained by Fourier-transform (FT) spectroscopy with about 0.01 cm$^{-1}$ accuracy covering about 85\% of the $B^1\Pi$ potential well. The corresponding point-wise IPA potential was constructed, which revealed a kink in the repulsive part of internuclear distance at about $R=4.1$~\AA ~attributed to avoided crossing of two $\Omega$ = 1 components belonging to $B^1\Pi$ and $c^3\Sigma^+$ states. A big variety of local perturbations have been observed, which could be useful to perform a global deperturbative analysis of the $B-b-c$ complex. The existing experiment-based $b^3\Pi$ state information is enough accurate and detailed~\cite{PRA2013}. Therefore, the key issue is the lack of reliable experiment-based information on the $c^3\Sigma^+$ state itself. Recently the situation became more favorable due to the first spectroscopic study of the $c^3\Sigma^+$ state of KCs applying the highly sensitive two-color polarization-labeling spectroscopy~\cite{Szczepkovski2017}. The authors have recorded rotationally-resolved spin-forbidden $c^3\Sigma^+ \leftarrow X^1\Sigma^+$ transitions with the accuracy of line positions better than 0.1 cm$^{-1}$. 646 rovibronic term values belonging to the $e$-component of the $c^3\Sigma^+(\Omega=1^+)$ state and spanning within vibrational quantum numbers $v^{\prime}$ from 5 to 29 and rotational quantum numbers $J^{\prime}$ from 38 to 106 were used to construct the IPA potential. The obtained empirical potential energy curve (PEC) has clearly demonstrated a non-monotonic shape, which was attributed to the pronounced SO interaction with the attractive part of the $b^3\Pi(\Omega=1)$ state near $R$ about 7~\AA. However, the already existing spectroscopic information is still not sufficient to realize a comprehensive deperturbation treatment of the $B^1\Pi - b^3\Pi - c^3\Sigma^+$ complex of the KCs molecule. It might be supposed that one of the reasons is the incompleteness of experimental information of the $c^3\Sigma^+$ state, in particular, for the $f$-component, as well as for the higher vibrational levels, which exhibit a strong impact of the intramolecular interaction with both $B^1\Pi$ and $b^3\Pi$ states. Furthermore, a realistic guess of the deperturbed potential energy curves and corresponding SO coupling functions are indispensably required in order to make the global deperturbation analysis of the $B-b-c$ complex a feasible procedure. Up to now, a high fidelity of the available \emph{ab initio} calculations accomplished on the states treated in non-relativistic~\cite{Habli2020}, scalar relativistic~\cite{Kim} and perturbative relativistic~\cite{Korek2006} approximations is still in  question.

These circumstances motivate a dual goal of the present study as a necessary step towards future enabling a full deperturbation of $A-B-b-c$ complex. The first goal was to extend the set of experimental term values of the $c^3\Sigma^+$ state given in~\cite{Szczepkovski2017} with increased accuracy of 0.01 cm$^{-1}$ or better with improved representation of both $e$- and $f$-components in the expanded range of $v_c$ and $J_c$ values. To do that we recorded and analyzed the FT spectra of triplet-triplet $c^3\Sigma^+ \rightarrow a^3\Sigma^+$ LIF induced by singlet-triplet $c \leftarrow X$ transitions. The second goal was to accomplish the relativistic multi-reference coupled-cluster calculation on the potential energy curves for the $B^1\Pi(\Omega = 1)$, $c^3\Sigma^+(\Omega = 0^-, 1)$ and $b^3\Pi(\Omega = 0^\pm, 1, 2)$ states, as well as on spin-forbidden $c^3\Sigma^+(\Omega = 1)  - X^1\Sigma^+(\Omega = 0^+)$ and spin-allowed $c - a$ transition dipole moments. The resulting relativistic adiabatic PECs (corresponding to pure ($c$) Hund's coupling case) have been then converted into their scalar-relativistic counterparts (corresponding to quasi-diabatic Hund's coupling case ($a$)), and effective SO functions corresponding to fully \emph{ab initio} analogs of the relevant empirical SO matrix elements  appearing in the reduced coupled-channel deperturbation analysis.

\section{Experiment}

\subsection{Experimental details}

Laser induced fluorescence (LIF) spectra of the $c^3\Sigma^+ \rightarrow a^3\Sigma^+$ transition ($c \rightarrow a$ in short), see Fig.~\ref{Fig1c3S}, were recorded with FT spectrometer IFS125-HR (Bruker) using InGaAs detector. Spectral resolution was set as 0.03 cm$^{-1}$. KCs molecules were produced in a linear heat-pipe loaded with 10 g K and 5 g Cs. The heat-pipe was heated to about 300$^o$C. For excitation the Ti:Saphire laser Equinox/SolsTis (MSquared) was operated within 13800 - 11800 cm$^{-1}$ range. Laser frequency was measured by wavemeter HighFinesse WS7.

\begin{figure}
\includegraphics[width=\columnwidth]{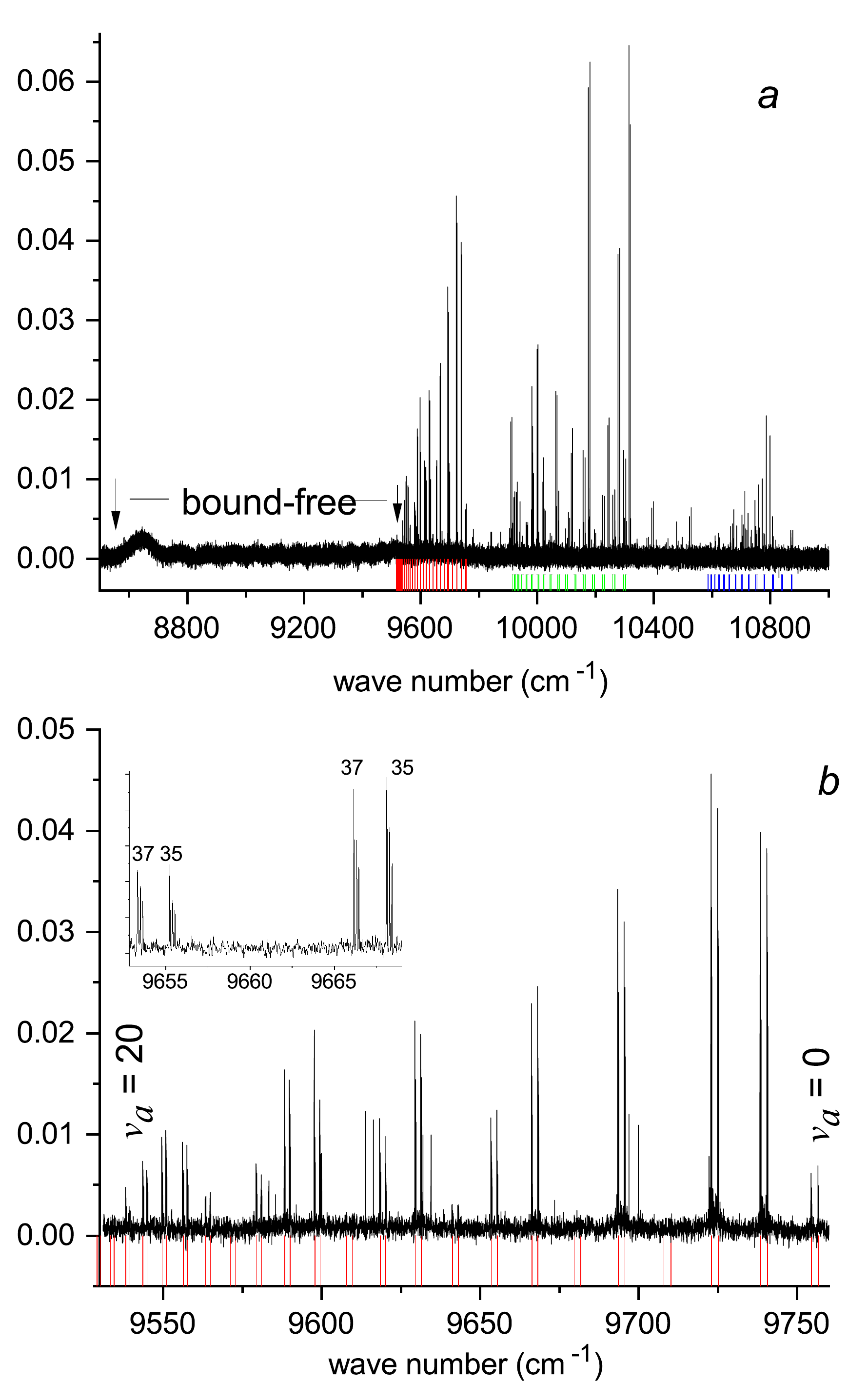}
\caption{KCs LIF spectrum recorded at laser frequency 13441.3065 cm$^{-1}$: (a) spectrum  overview. The high frequency part is cut off by a long-pass 900 nm edge filter. Long red bars below the spectrum mark the $c^3\Sigma^+ \rightarrow a^3\Sigma^+$ system. Short green and blue bars mark the two assigned $A^1\Sigma^+_{u} \rightarrow X^1\Sigma^+_{g}$ LIF progressions of the K$_2$ molecule. The low frequency part contains $c \rightarrow a$ bound-free LIF transitions with the last maximum at about 8600 cm$^{-1}$; (b) zoomed part of the spectrum in (a),  with a doublet $P$, $R$ progression from the $c$-state $e$-level with  $v_c$ = 23, $J_c$ = 36, $E_c$ = 13586.065 cm$^{-1}$. Here $v_a$  denotes the $a$-state vibrational level. The inset shows two doublets with a clearly seen hyperfine structure of the lines; numbers at the lines denote a rotational $N_a$ level of the $a^3\Sigma^+$ state.} \label{Fig2}
\end{figure}

In order to excite the $c^3\Sigma^+$ state from the ground $X^1\Sigma^+$ state three different approaches were applied. First, we calculated the excitation frequencies for the $c^3\Sigma^+ \leftarrow X^1\Sigma^+$ transitions, as well as the Franck-Condon factors (FCF) using the $c^3\Sigma^+$ effective empirical potential obtained in ~\cite{Szczepkovski2017} and the $X^1\Sigma^+$ potential from ~\cite{Ferber2013}. It appeared that the calculated excitation frequencies for rovibronic transitions $v_c, J_c \leftarrow v_X, J_X$ selected according to FCFs did not, as a rule, reproduce the targeted transition precisely enough, that is, within Doppler width. Therefore, the laser frequency was scanned, within some interval, around a calculated one and the LIF signal was monitored in the spectral range from 9000 to 10000 cm$^{-1}$ in Preview Mode of the spectrometer, which exposed the LIF spectrum in real time at low resolution. The laser frequency was fixed at the maximal value of the LIF signal. In order to excite vibrational levels higher than $v_c$ = 29 we started the experiment with highest accessible laser frequency and then decreased it step by step to excite lower vibrational levels. Usually, the spectra contained one strong and several substantially weaker $c \rightarrow a$ progressions from different $v_c$, $J_c$. Generally, along with $c \rightarrow a$ progressions. strong LIF transitions to the ground $X^1\Sigma^+$ state from the accidentally excited $A^1\Sigma^+$ and $b^3\Pi$ states of KCs, K$_2$, or Cs$_2$ molecules were also observed. Fortunately, at higher excitation frequencies these transitions were well separated from the $c \rightarrow a$ system, see Fig.~\ref{Fig2}. In order to eliminate these, sometimes very strong, LIF progressions, the long-pass filters FEL900, FEL 950, or FEL1000 were used, which cut off LIF above 11000, 10500, and 10000 cm$^{-1}$, respectively.

\begin{figure}
\includegraphics[width=\columnwidth]{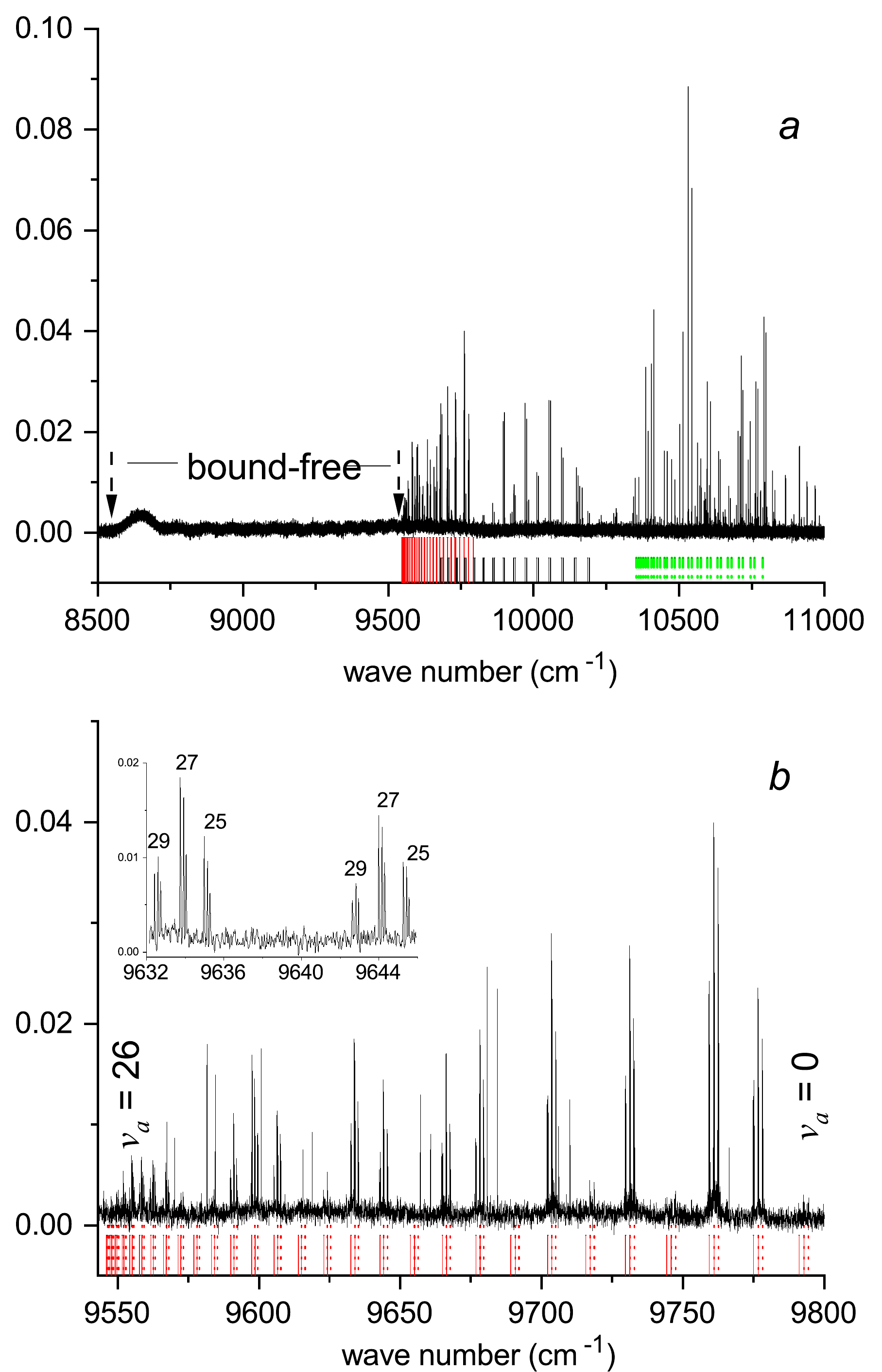}
\caption{KCs LIF spectrum recorded at laser frequency 13489.6223 cm$^{-1}$: (a) - spectrum  overview. The high frequency part is cut off by a long-pass edge 900 nm filter. Long bars below the spectrum mark $c^3\Sigma^+ \rightarrow a^3\Sigma^+$ system. Short bars mark two assigned  $A^1\Sigma^+_{u} \rightarrow X^1\Sigma^+_{g}$ LIF progressions of K$_2$ molecule. The low frequency part contains bound-free LIF transitions with last maximum at about 8600 cm$^{-1}$; (b) - zoomed part of the spectrum in (a) with  $P$, $Q$ and $R$ progression from the $c$-state $f$-level with $v_c$ = 21, $J_c$ = 27, $E_c$ = 13614.689 cm$^{-1}$. The inset shows $P^{Q}$, $Q$, $R^{Q}$ lines with hyperfine structure.} \label{Fig3}
\end{figure}

Approaching lower vibrational $v_c$ levels, weak $c \rightarrow a$ systems start to partially overlap with stronger $A \rightarrow X$ systems, therefore the Preview Mode approach was practically not applicable for $v_c$ below 13. To overcome this difficulty, we filtered out the whole discrete spectrum and, during scanning the laser frequency, monitored the emergence of the last maximum at about 8600 cm$^{-1}$ in a continuous bound-free part of the $c \rightarrow a$ progressions, which almost always took place when the $c$-state was excited, see Figs.~\ref{Fig2},~\ref{Fig3}.

Below $v_{c}$ = 5 the last maximum of bound-free transitions was practically unobservable because of substantially decreased intensity of the $c \rightarrow a$ LIF. Since there was no chance to reproduce the term values of these low levels precisely enough using the PEC from~\cite{Szczepkovski2017}, a preliminary potential was constructed, which combined the present data with the data from ~\cite{Szczepkovski2017}. A critical comparison of present measured term values with the ones calculated by the new potential allowed us, after several trials, to predict the correct excitation frequency and to observe $v_c$ = 4. By repeating this 'new data - new potential' approach several times, it was possible to observe even the lowest $v_c$ = 0 level. Note that the favorable FCFs of the $c - X$ system for excitation of lower $v_{c}$ levels are shifting to higher $v_X$; e. g. $v_X$ = 13 was used to excite $v_c$ = 0.

\subsection{Spectra analysis and assignment}

According to full relativistic electronic structure calculation, see Section~\ref{FSRCC}, the $c^3\Sigma^+$ state is the inverted $\lambda_c = E(\Omega=1)-E(\Omega=0^-)< 0$ triplet corresponding to the intermediate $(a)\leftrightarrow (c)$ Hund's coupling case, while the lowest triplet $a^3\Sigma^+ (\lambda_a \approx 0)$ state corresponds to the pure $(b)$ Hund's case~\cite{Watson1971}. The direct excitation of the fine structure $c^3\Sigma^+_{\Omega=0^-}$ component from the ground $X^1\Sigma^+(X0^+)$ state is not possible since the $c0^--X0^+$ transition is strictly forbidden even in pure $(c)$ Hund's case. At the same time, its $c1^{\pm}$ components with $1^+$ and $1^-$ being the $e$- and $f$-components, respectively, can be excited due to the strong SO coupling of the $c^3\Sigma^+_{\Omega=1^{\pm}}$ components with the nearby $B^1\Pi$ state (see, for instance, Fig.~\ref{Fig1c3S}).

As a result, two different kinds of $c^3\Sigma^+ (v_c, J_c) \to a^3\Sigma^+(v_a, N_a)$ LIF spectra were recorded; here $N_a$ denotes the rotational quantum number of the  $a^3\Sigma^+$ state in Hund's case ($b$).  Most frequently the doublet progressions consisting of $P$, $R$ ($N_a=J_c\pm 1$) branches were observed. However, in a number of cases progressions with three branches $P^{Q}$, $R^{Q}$, and $Q$, where $N_a=J_c^{\prime}+2$ for $P^{Q}$, $N_a=J_c^{\prime}-2$ for $R^{Q}$, and $N_a=J_c^{\prime}$ for $Q$, respectively, have been recorded. The observed doublet progressions originate from the rovibronic $e$-levels solely belonging to the $\Omega=1^+$ component of the $c^3\Sigma^+$ state, while the triplet progressions originate from the $f$-levels of the $\Omega=1^-$ component, which are rotationally mixed with the close-lying $\Omega=0^-$ component of the same $c$-state. The $\Omega=1^+$ and $\Omega=1^-$ components of the $c^3\Sigma^+$ state are excited from the ground $X^1\Sigma^+$ state according to $P$, $R$ ($J_c=J_X\pm 1$) and $Q$ ($J_c=J_X$) transitions, respectively.

An example of the doublet $P, R$-progression is shown in Fig.~\ref{Fig2}. The discrete part of the $c \rightarrow a$ system appears within the range from 9500 to 10000 cm$^{-1}$ and is marked by vertical lines below the spectrum. The lower frequency range of the same LIF progression contains a continuous bound-free part of the $c\to a$ transition with oscillating intensity; with a pronounced last maximum at about 8600 cm$^{-1}$. The  high frequency range above 9800 cm$^{-1}$ contains the $A^1\Sigma^+\sim b^3\Pi \to X^1\Sigma^+$ doublet progressions of K$_2$. The discrete $c\rightarrow a$ part is zoomed in Fig.~\ref{Fig2}b. Note that due to the hyperfine structure (HFS) in the $a^3\Sigma^+$ state each rotational line is split into three groups of unresolved lines, each of them containing a large number of HFS transitions~\cite{Ferber2009}, see inset in Fig.~\ref{Fig2}b. Assignment of the $c \rightarrow a$ progressions was based on the accurate empirical PEC of the $a^3\Sigma^+$ state obtained in~\cite{Ferber2013}. Only the central HFS group was used for assignment since it is expected to be less affected by the hyperfine interaction; its position was used for the $a$-state PEC construction in~\cite{Ferber2013,Ferber2009}. The rovibronic term values $E_c(v_c, J_c)$ of the $c$-state were obtained by adding the energy $E_a$($v_a$, $N_a$) of the lower $a^3\Sigma^+$ state to the frequency of the respective transition. The uncertainty of the obtained term values is about 0.01 cm$^{-1}$ because the width of an absorption transition is about 0.015 cm$^{-1}$ due to Doppler effect. In a number of spectra, around the strong lines, the weaker satellite lines appearing due to collisional population transfer to neighbouring rotational levels in the upper state were recorded and assigned. All energies obtained for the $c$-state were uniformly shifted down by 0.035 cm$^{-1}$ in order to refer to the ground $X$-state. This correction was additionally checked at the early stage of the experiment by recording of a very weak $c\rightarrow X$ LIF transitions originating from the same level of the $c$-state.

An example of LIF spectrum recorded at excitation of the 1$^-$ state $f$-level in $Q$ transition is presented in Fig.~\ref{Fig3}. Generally, the recorded spectrum, see Fig.~\ref{Fig3}a, is similar to the one in Fig.~\ref{Fig2}a. The discrete part of the $c \rightarrow a$ system is shown in Fig.~\ref{Fig3}b. The observed progression consisting of three branches is clearly seen in the inset of Fig.~\ref{Fig3}b. An attempt to assign them separately as two doublet progressions yields to the two upper state rotational levels $J_c$ = 26 and 28 with the same energy $E_c$, which is impossible. Hence, the upper state is $J_c$ = 27 and the transitions to rotational levels with $N_a=25 (R^{Q})$, 27$(Q)$, and 29$(P^{Q})$ take place.

\begin{figure}
\includegraphics[width=\columnwidth]{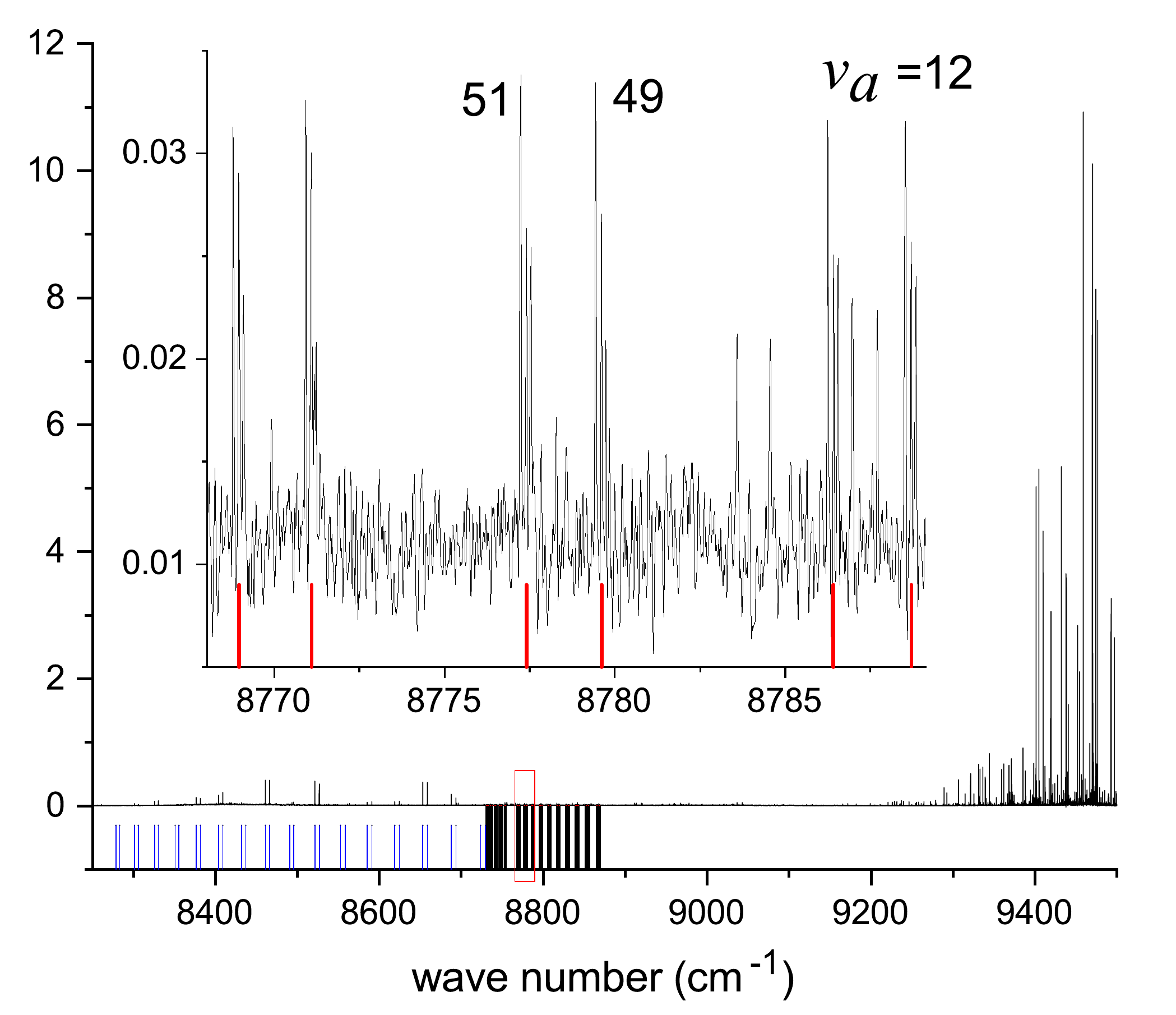}
\caption{KCs LIF spectrum recorded at laser frequency 11824.2221 cm$^{-1}$. Long bars below the spectrum mark the range where a weak $c \rightarrow a$ LIF progression from the level $v_c$ = 0, $J_c$ = 50, $E_c$ = 12788.846 cm$^{-1}$ was recorded, see the  fragment for $v_a$ = 12-14 in the inset. Short bars in the lower frequency part mark a KCs $A \sim b \rightarrow X$ progression, while the higher frequency part contains many K$_2$ $A^1\Sigma^+_{u} \rightarrow X^1\Sigma^+_{g}$ progressions.} \label{Fig4J50}
\end{figure}

An example of LIF spectrum originating from $v_c$ = 0 is given in Fig.~\ref{Fig4J50}. As seen, the $c \rightarrow a$ progression is extremely weak. The progression was recorded at the highest possible amplification and with increased acquisition time.

It should be noted that the correctness of assignment of $E_c$ and $J_c$ was always checked by the coincidence of calculated absorption $c(J_c,E_c) \leftarrow X(v_X,J_X)$ frequency with the frequency of laser excitation.

\section{Theory}

\subsection{Relativistic electronic structure calculation}\label{FSRCC}

The fully relativistic coupled cluster calculation of the potential energy curves, SO coupling functions and transition dipole moments among several low-lying states of KCs has been performed in the present work in order to facilitate a spectroscopic analysis of the $c$-state.

The employed scheme of excited state calculations generally resembles that described and used in Refs.~\cite{Zaitsevskii:17,Krumins:20}. The relativistic electronic structure model was defined by the accurate semi-local shape-consistent two-component pseudo-potential of the ``small'' atomic cores ($1\!-\!2s,\,2p$ K, $1\!-\!4s,\,2\!-\!4p,\,3\!-\!4d$  Cs), derived from the valence-shell solutions
of the atomic Dirac--Fock--Breit equations with the Fermi nuclear charge model~\cite{Mosyagin:10a,ourGRECP}. The correlations of 18 valence and sub-valence electrons were treated explicitly using the modified version~\cite{Zaitsevskii:17} of the Fock space relativistic coupled cluster (FS-RCC) method~\cite{Eliav:98,Visscher:01}. We started with building the molecular spinors and the Fermi vacuum state by solving the spin-orbit-coupled SCF equation for the ground state of KCs$^{2+}$. Contracted Gaussian basis sets were used to expand the components of one-electron spinors. The $[7s\,7p\,6d\,4f\,3g\,1h]$ basis for Cs was taken from Ref.~\cite{Zaitsevskii:17}; a $[7s\,7p\,6d\,4f\,2g]$ basis set for K was built using the same principles. The complete model spaces in the one- and two-particle Fock space sectors corresponding to KCs$^+$ and neutral KCs were defined by 61 Kramers pairs of 'active' lowest virtual spinors. The cluster operator expansion comprised only single and double excitations (FS-RCCSD approximation).

Numerical instabilities, which could arise from the presence of numerous intruder states were suppressed by using the simulated imaginary shift technique~\cite{Oleynichenko:20cpl}. We assumed the uniform shift amplitude value $s_{K_2} = s_2 = -0.6$ (see Eq.(8) in Ref.~\cite{Oleynichenko:20cpl}) for all double excitations $K_2$ destroying two valence particles, and $s_{K_1}=s_2/2$ for the excitations $K_1$ affecting one valence particle. The attenuation parameter $m=3$ was chosen to ensure stability of iterative procedure of amplitude equations solution. The calculations for the Fermi vacuum sector were always performed with  non-shifted energy denominators; this was essential for ensuring the exact core separability of the results.

The FS-RCC calculations were performed using the appropriately modified DIRAC17~\cite{DIRAC:17,DIRAC:20} and EXP-T~\cite{EXPT:20} program packages; in the latter case, DIRAC17 still was used to solve SCF equations and to transform the molecular integrals.

The potential energy curves for the adiabatic excited states were obtained by adding the vertical FS-RCC excitation energies as functions of the internuclear separation to the highly accurate empirical ground-state potential~\cite{Ferber2013,Ferber2009} (cf. Refs.~\cite{Pazyuk:15,Zaitsevskii:17}). The resulting curves were converted into quasi-diabatic potentials and effective SO interaction functions (\emph{ab initio} analogs of the corresponding empirical functions appearing in local deperturbation analysis) through projecting the scalar relativistic eigenstates on the subspace of strongly coupled eigenstates of the total Hamiltonian~\cite{Zaitsevskii:17}. At this stage the many-electron wave functions were approximated by their projections onto the FS-RCC model space.

Transition electric dipole moments were evaluated using the finite-field scheme~\cite{Zaitsevskii:18}. Although the calculations also involved only the model space parts of many-electron wavefunctions, the resulting transition moment values implicitly incorporated the bulk of the contributions from the remainder part of these wavefunctions~\cite{Zaitsevskii:98,Zaitsevskii:20}. To prevent the deviations of resulting transition moment matrix from exact Hermiticity, we performed the preliminary transformation of the non-Hermitian FS-RCC effective Hamiltonians to their Hermitian counterparts \emph{via} the symmetric orthogonalization of their eigenvectors.

\subsection{The empirical potential construction accompanied by the $\Omega=1^{\pm}$-doubling treatment}\label{DPF}

The experimental rovibronic term values currently assigned to both $e/f$-components of the $c^3\Sigma^+_{1^{\pm}}$ state have been simultaneously treated within a direct-potential-fit (DPF) analysis with a twofold aim: to refine the interatomic potential and to elucidate the origin of the pronounced $\Omega$-doubling effect in the $\Omega=1^{\pm}$ component of the triplet $c^3\Sigma^+$.

The empirical point-wise IPA potential $U_c^{IPA}(R)$ and the analytical parameters of the $\Omega$-doubling functions $q_c(R)$ were determined for the $c^3\Sigma^+(\Omega=1^{\pm})$ state during the iterative minimization of the non-linear functional:
\begin{eqnarray}\label{chisquared}
\chi^2=\sum_{i=1}^{N_{expt}}\frac{(\delta^{expt}_i)^2}{(\sigma^{expt}_i)^2+(\delta^{expt}_i)^2/3}+w\sum_{i=1}^{N_{ab}}\left (\frac{\delta^{ab}_i}{\sigma^{ab}_i}\right)^2 ,\\
\delta^{expt}_i=E^{expt}_i-E^{calc}_i;\quad \delta^{ab}_i=U_c^{ab}(R_i)-U_c^{IPA}(R_i) ,\nonumber
\end{eqnarray}
where the difference-based relativistic PEC $U_c^{ab}(R)$ constructed in Section~\ref{FSRCC} was inserted in Eq.(\ref{chisquared}) with an appropriate weight $w$ to propagate smoothly the empirical potential outside the experimental data region. The uncertainties of the present term values $\sigma^{expt}_i$ were taken as equal to 0.01 cm$^{-1}$. The previous experimental data set from~\cite{Szczepkovski2017} was also involved in the DPF with $\sigma^{expt}_i=0.05$ cm$^{-1}$. The required uncertainties $\sigma^{ab}_i$ in the \emph{ab initio} PEC $U_c^{ab}(R_i)$ were roughly estimated by a comparison with its previous theoretical counterparts.

The theoretical term values $E^{calc}$ were obtained from the numerical solution of the radial equation:
\begin{eqnarray}\label{radial}
\left(\frac{\hbar^2 d^2}{2\mu dR^2} + U_c^{IPA} + B[1 + sBq_c]X- E^{calc}\right)|v_c^J\rangle=0;\\
B(R)\equiv \frac{\hbar^2}{2\mu R^2};\qquad X\equiv J(J+1)-\Omega^2, \nonumber
\end{eqnarray}
where $s$ is the flag switch from 0 or 1 for the $e$ or $f$ parity levels, respectively.

The expectation value of $q_c(R)$ function introduced in Eq.(\ref{radial}) as
\begin{eqnarray}\label{qfactor}
q_{\Omega}(v_c,J_c)\approx \langle v^J_c|q_c B^2|v^J_c\rangle
\end{eqnarray}
determines the so-called $q_{\Omega}$-factor, which characterises the $\Omega$-doubling effect~\cite{Field2004book}
\begin{eqnarray}\label{qfactor1}
E_{\Omega=1^+}^e - E_{\Omega=1^-}^f = q_{\Omega}[J(J+1)-1].
\end{eqnarray}

The corresponding $q_c(R)$ function of the KCs $c(\Omega=1)$-state has been defined empirically as the quadratic polynomial of the reduced coordinate $y\in [-1,+1]$:
\begin{equation} \label{Qfunction}
q_c(y)=q_0+q_1y+q_2y^2;\quad  y=\frac{R/R_{ref}-1}{R/R_{ref}+1},
\end{equation}
where $R_{ref}=5.2$~\AA~is the fixed reference distance, while $q_i$ are the fitting coefficients.

It should be noticed that the origin of the $\Omega$-doubling effect in the triplet $^3\Sigma^{+}$ states~\cite{Watson1971} completely differs from a nature of the conventional $\Lambda$-doubling effect observed in the $^{1,3}\Pi$ states~\cite{NaK1998}. The latter is induced by the regular electronic-rotational interaction (so called $L$-uncoupling or Coriolis effect) with the remote $^{1,3}\Sigma^{+}$ states, while the first one is caused by the local spin-rotational interaction with the nearby $\Omega=0^-$-component belonging to the same $^3\Sigma^{+}$ state:
\begin{eqnarray}\label{Qfunction1}
q_{\Omega=1} = -\frac{2\eta^2}{\lambda};\quad \eta \approx \sqrt{S(S+1)-\Sigma(\Sigma\pm 1)},
\end{eqnarray}
where $\lambda=E_{\Omega=1}-E_{\Omega=0^-}$ is the splitting of the $\Omega=1$ and $\Omega=0^{-}$ components of the $^3\Sigma^{+}$ state that is mainly determined by the SO interaction with the singlet $^1\Pi$ states manifold:
\begin{eqnarray}\label{lambda}
\lambda \approx \sum_{^1\Pi}\frac{|\xi^{so}_{^1\Pi-^3\Sigma^+}|^2}{E^{^3\Sigma^+}-E^{^1\Pi}}.
\end{eqnarray}
If the molecular spin $S$ is conserved, then $\eta \approx \sqrt{2}$ for the $^3\Sigma^{+}$ state (where $S=1$ and its projection $\Sigma=0$). Thus, the $q_{\Omega}^{^3\Sigma^{+}}$-values should be expected to be significantly larger than their $q_{\Omega}^{^{1,3}\Pi}$ counterparts since the $\lambda$-splitting of $^3\Sigma^{+}$ states is normally much smaller than the energy distance between the $\Pi$ and $\Sigma^{+}$ states of the same multiplicity.

\subsection{Radiative properties estimate}\label{intensity}

\emph{Ab initio} spin-allowed and spin-forbidden electronic transition dipole moment (TDM) functions $d^{ab}_{cj}(R)$ evaluated in Sect.~\ref{FSRCC} between the $c^3\Sigma^+$ state and lower-lying $j\in [a^3\Sigma^+, b^3\Pi, X^1\Sigma^+]$ states (see Fig.~\ref{Fig1c3S}) were applied to estimate the radiative lifetimes $\tau_c$ and vibronic branching ratios $R_{c\to j}$ of the rovibrational levels of the $c$-state observed in the present work~\cite{TellinghuisenCPL84,PupyshevCPL94}:
\begin{eqnarray}\label{tausum}
\frac{1}{\tau_c}&=&k\langle v_c^J|\sum_j [\Delta U^{ab}_{cj}]^3[d_{cj}^{ab}]^2|v_c^J\rangle,\\
R_{c\to j}&=& \tau_c \times \left [k\langle v_c^J|[\Delta U^{ab}_{cj}]^3[d^{ab}_{cj}]^2|v_c^J\rangle \right ],
\end{eqnarray}
where $\Delta U^{ab}_{cj}=U^{ab}_c(R)-U^{ab}_j(R)$ is the difference of the \emph{ab initio} PECs and $k\equiv 8 {\pi }^2/3\hbar {\epsilon_0} = 2.0261\times 10^{-6}$ if energies are given in cm$^{-1}$ and electronic TDM  values in a.u. The rovibrational wave functions $|v_c^J\rangle$ required for the upper $c$-state were obtained from the solution of the radial equation (\ref{radial}) with the IPA potential constructed in Sect.~\ref{DPF}.

\section{Results and Discussion}
\subsection{Experimental term values and the empirical PEC}

The presently obtained $c^3\Sigma^+$ state rovibronic term values are depicted in Fig.~\ref{Fig5DF} and presented in Table I of Supplementary material. For visibility, the term values for the first two vibrational levels of the $B^1\Pi$ state are also presented in the figure. One can see that most of the obtained $c$-state term values are well below the $B$-state. Overall we have determined 127 $f$-levels and 539 $e$-levels covering $v_c$-range from $v_c$ = 0 to about 44. It should be mentioned that for the $v_c>34$ it might be difficult to assign unambiguously the levels to the $c$- or $B$-state because of their strong mutual SO mixing; a more detailed analysis would be necessary. As seen, the range of the currently observed rotational quantum numbers $J_c$ spreads within the values from $J_c$ = 11 to 149. In seven LIF spectra the $P, R$ progressions belonging to the $^{41}$KCs isotopologue were observed and assigned.

\begin{figure}
\includegraphics[width=\columnwidth]{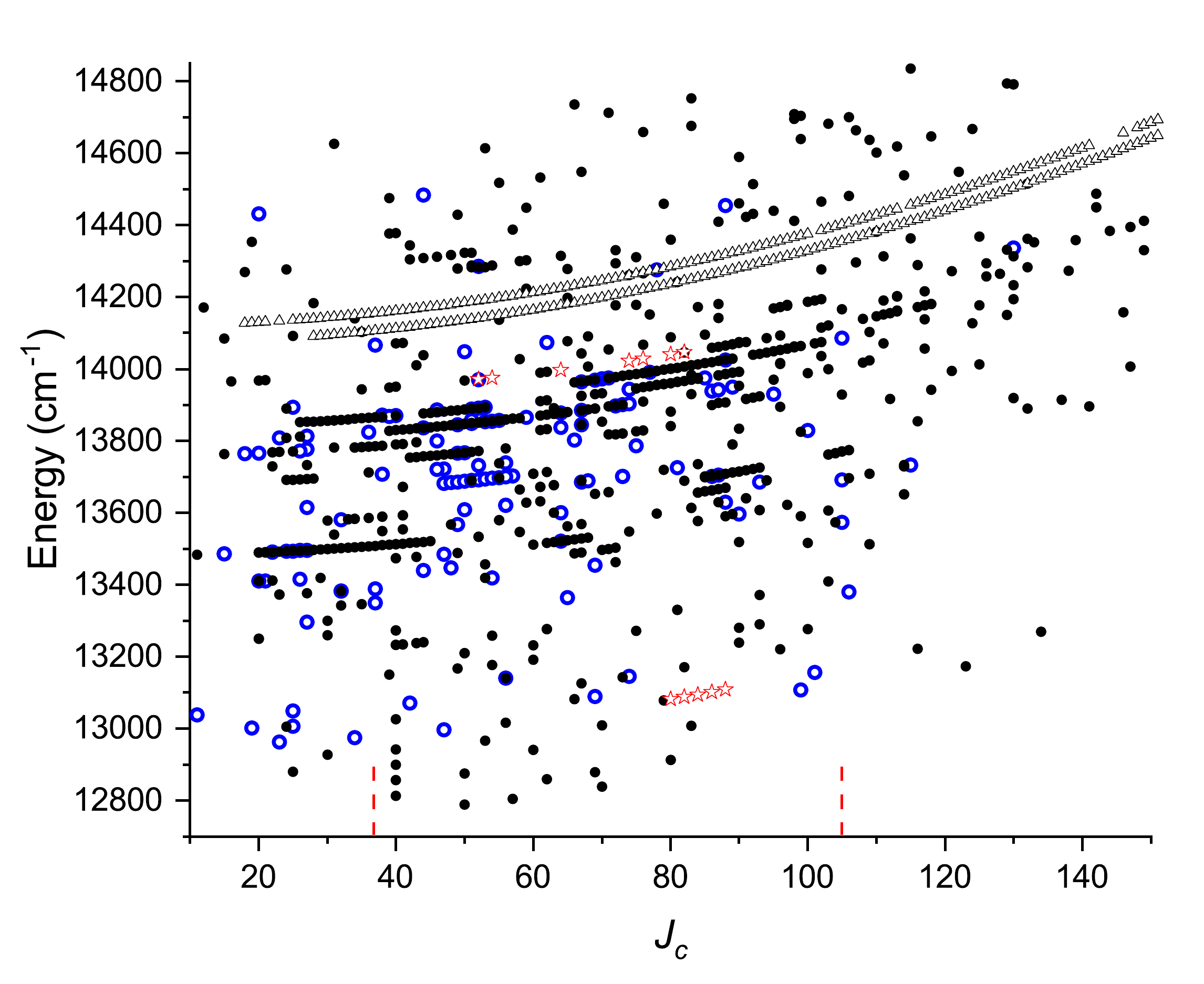}
\caption{Distribution of present rovibronic term values of the  $c^3\Sigma^+_{1}$ state as dependent on $J_c$: dots - $e$-levels, empty circles - $f$-levels. Empty red asterisks mark the lowest $v_c$ = 5  and the highest $v_c$ = 29 vibrational levels measured in \cite{Szczepkovski2017}; dashed vertical lines mark $J_c$ = 38 - 106 range in Ref.~\cite{Szczepkovski2017}.
Triangles are the term values of the $B^1\Pi$ state for $v_B$ = 0, 1 measured in Ref.~\cite{Birzniece2012}.} \label{Fig5DF}
\end{figure}

\begin{figure}
\includegraphics[width=\columnwidth]{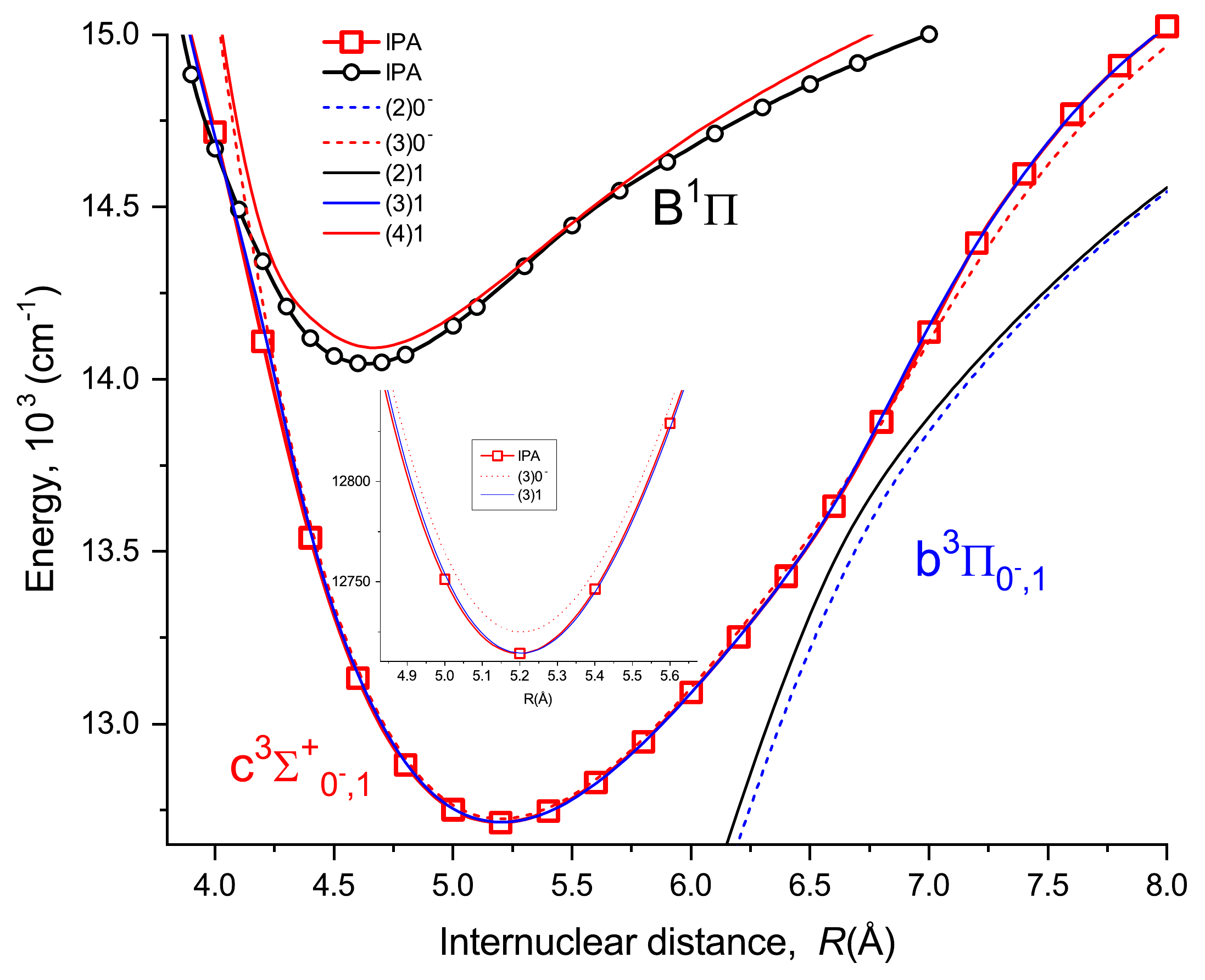}
\caption{Empirical (IPA) and \emph{ab initio} (FS-RCC) potential energy curves obtained for the relativistic (adiabatic) electronic states constituting the $B^1\Pi \sim b^3\Pi \sim c^3\Sigma^+$ complex of KCs. The IPA potential for the $B^1\Pi$ state was taken from Ref.\cite{Birzniece2015}. The inset demonstrates the inverted order of the \emph{ab initio} $\Omega=1$ and $\Omega=0$ components of the $c^3\Sigma^+$ state. All original \emph{ab initio} PECs were uniformly downshifted by about 70 cm$^{-1}$ to match smoothly the minimum of the present IPA potential for the $c^3\Sigma^+_1$ state.}\label{Fig2c3S}
\end{figure}

To construct the effective empirical interatomic PEC that would describe the present term values of the $c$-state with an accuracy close to the experimental one, we included in the least squares fit (\ref{chisquared}) only the data for vibrational levels from $v_c$ = 0 till 22, which are all located below the energy region of avoided crossing with $\Omega$ = 1 component of the $b(1)^3\Pi$ state, see Fig.~\ref{Fig2c3S}. These data were supplemented with the data from~\cite{Szczepkovski2017} for $v_c\in [5,22]$. Overall the fitted values contain 201 $e$-level and 54 $f$-levels obtained in present work and 469 $e$-levels borrowed from~\cite{Szczepkovski2017}. It should be noted that our attempts to extend the energy range of the IPA potential by including in the fit the higher vibrational levels had failed to reproduce the $v$ below 22 levels with a required spectroscopic accuracy.

\begin{figure}
\includegraphics[width=\columnwidth]{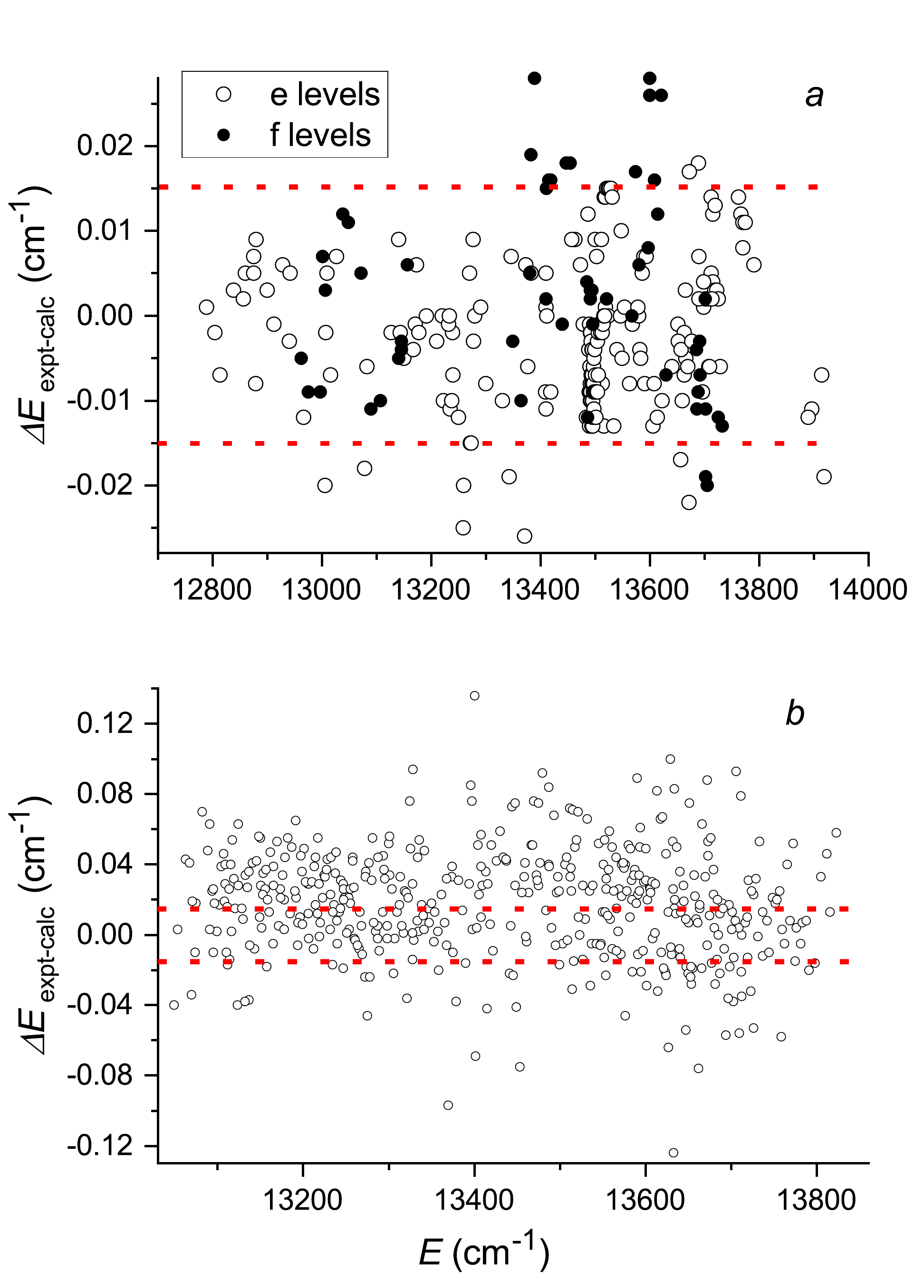}
\caption{Residuals $E_{expt} - E_{IPA}$ for $c$-state rovibronic levels measured in the present experiment (a) and in Ref.~\cite{Szczepkovski2017} (b).} \label{Fig6resAS5}
\end{figure}

The resulting IPA potential for the $c^3\Sigma^+_1$ state of KCs is depicted in Fig.~\ref{Fig2c3S} and presented in Supplementary material, see Table II. It includes 24 grid points within $R$-range from 3.4 to 8.0~\AA,  which is minimal necessary to reproduce the level energies with the experimental accuracy. The fitted coefficients of the corresponding $\Omega$-doubling function $q_c(R)$ are given in Table~\ref{tablq}. The residuals between the present experimental term values and the respective values given by the fit are presented in Fig.~\ref{Fig6resAS5}a. As seen, about 90\% of points are within $\pm$0.015 cm$^{-1}$ corridor. The residuals for the data from~\cite{Szczepkovski2017} are presented in Fig.~\ref{Fig6resAS5}b. The residuals, along with fitted term values, as well as the fitted $\Omega$-splitting constant $q_{\Omega}$ values are presented in Table II of Supplementary material.

To confirm the mass-invariant properties of the derived IPA potential we evaluated the term value for the $v_c$ = 17; $J_c$ = 42 $e$-level of the less abundant $^{41}$KCs isotopologue by according changing the reduced molecular mass $\mu$ in the radial equation (\ref{radial}). The calculated energy 13461.927 cm$^{-1}$ coincides with its experimental counterpart within 0.016 cm$^{-1}$ .

\begin{table}
\caption{The empirical (fitted) coefficients (in 1/cm$^{-1}$) of the quadratic polynomial (\ref{Qfunction}) describing the observed $\Omega$-splitting of the $c^3\Sigma^+_1$ state of KCs.} \label{tablq}
\begin{center}
\begin{tabular}{cc}
\hline\hline
$q_0$     &  +0.29977\\
$q_1$     &  -1.99781\\
$q_2$     &  -25.9410\\
\hline
\end{tabular}
\end{center}
\end{table}

\subsection{Relative intensity distributions}

It is well enough established that a direct comparison of experimental relative intensity distributions in long (ideally full) LIF progressions with their theoretical counterparts is a critical independent test of a correctness of the PECs available for both upper and lower optically connected electronic states. Therefore, we have simulated intensity distributions for the several observed $c\to a$ LIF progressions of KCs in order to additionally test the IPA potential derived for the upper $c$-state tacitely assuming that the lower $a$-state PEC is perfectly known at least in its bound region~\cite{Ferber2013,Ferber2009}.

The required theoretical intensities $I_{c\to a}^{calc}$ were calculated as
\begin{eqnarray}\label{Iten}
I_{c\to a}^{calc}(v_c,v_a) \sim \nu^4_{ca}|\langle v_c^J|d^{ab}_{ca}|v_a^N\rangle|^2\;\qquad \nu_{ca}=E_{v_c^J}-E_{v_a^N},
\end{eqnarray}
where rovibronic eigenvalues $E_{v_c^J}$ and eigenfunctions $|v_c^J\rangle$ of the $c$-state were obtained by the solution of radial equation (\ref{radial}) with the present IPA potential. The empirical $a$-state potential $U^{emp}_a$ was borrowed from Ref.~\cite{Ferber2013} to calculate the corresponding energies $E_{v_a^N}$ and wavefunctions $|v_a^N\rangle$ of the $a$-state.

\begin{figure}
\includegraphics[width=\columnwidth]{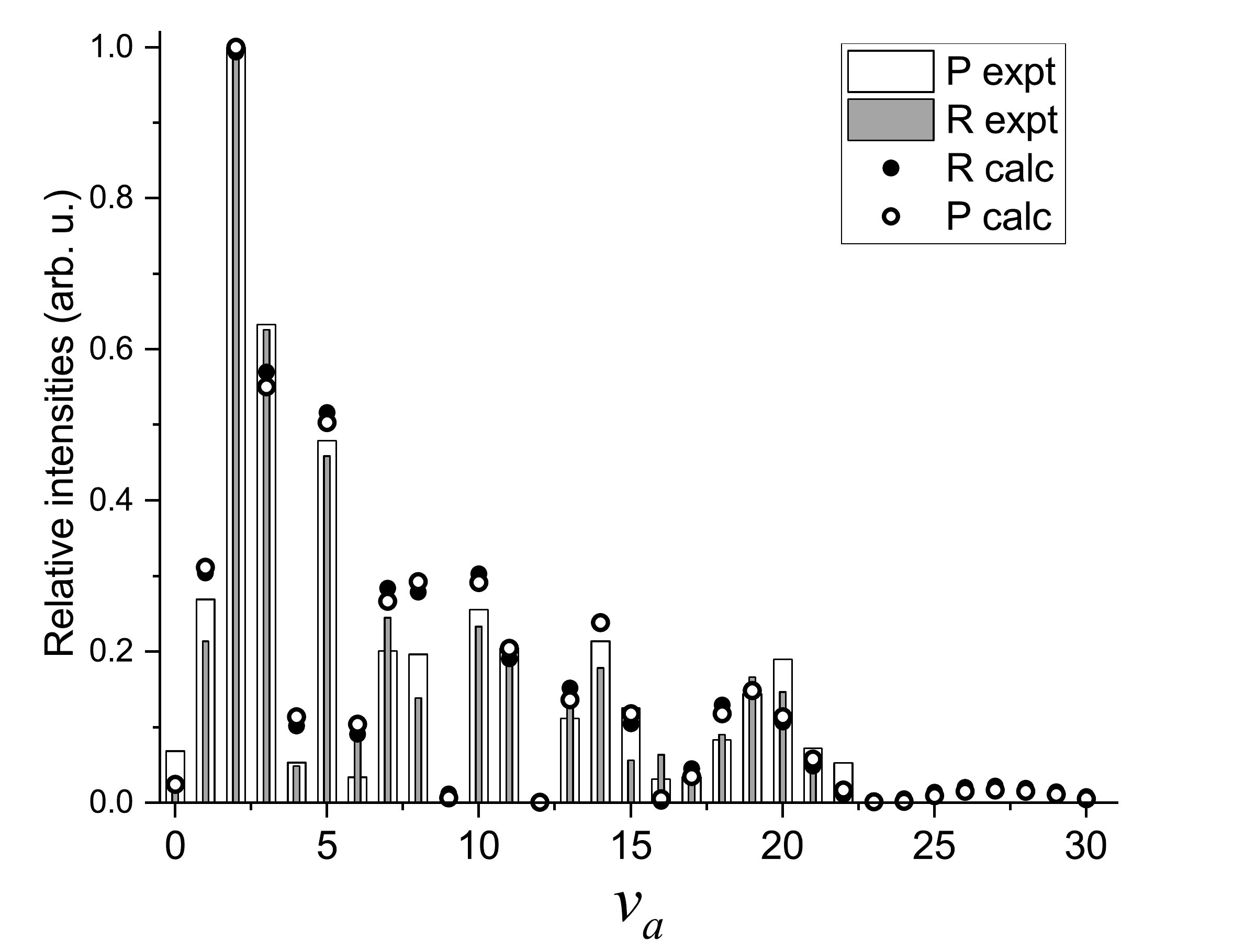}
\caption{Experimental $I^{expt}_{c\to a}$ and calculated $I^{calc}_{c\to a}$ relative intensity distributions in the $c\to a$ LIF progression from the $v_c$ = 23, $J_c$ = 36 upper level.} \label{Fig7}
\end{figure}

\begin{figure}
\includegraphics[width=\columnwidth]{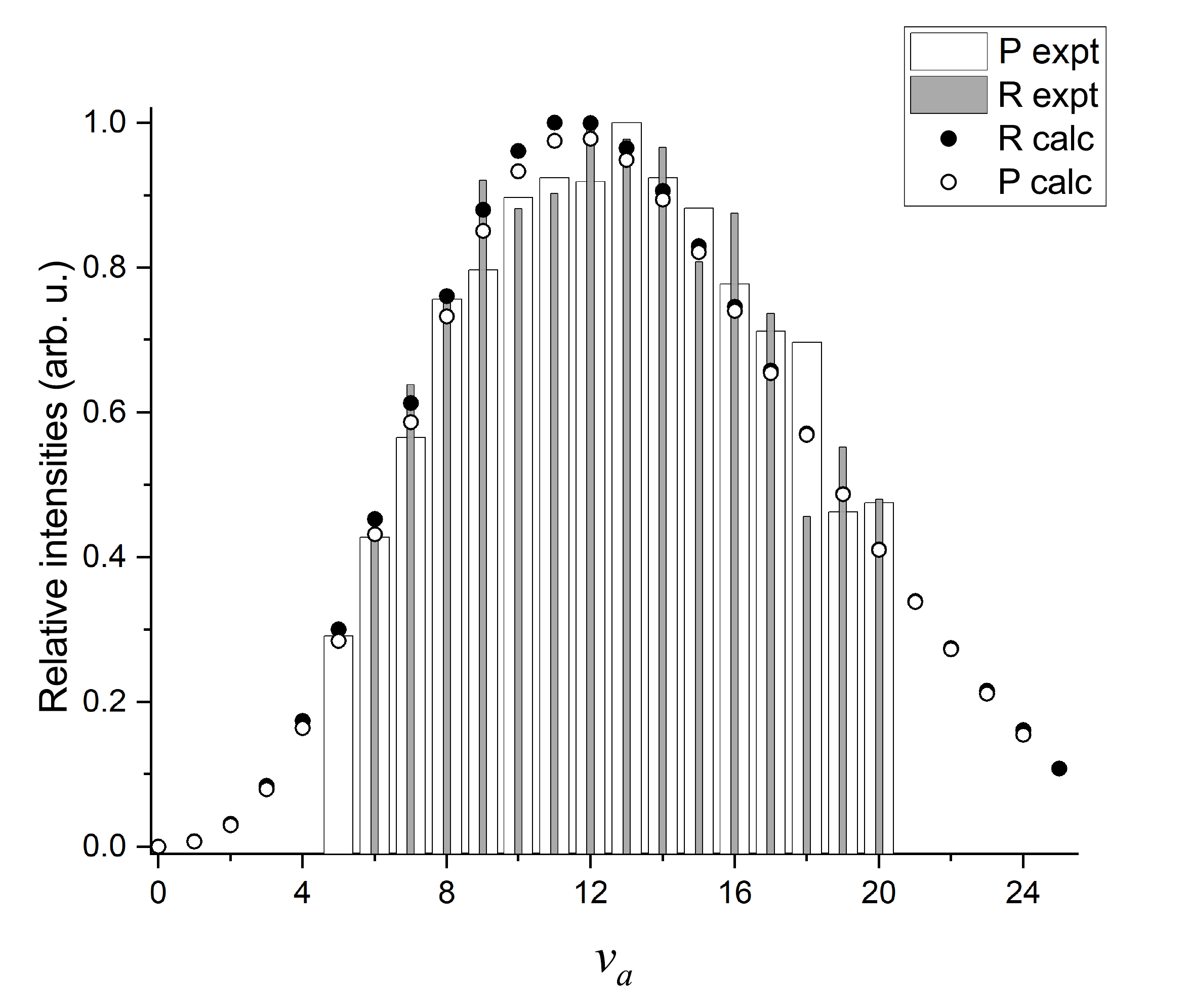}
\caption{Experimental $I^{expt}_{c\to a}$ and calculated $I^{calc}_{c\to a}$ relative intensity distributions in the $c\to a$ LIF progressions from the $v_c$ = 0, $J_c$ = 50 upper level.} \label{Fig7intJ50}
\end{figure}

The relative LIF intensity distribution have been measured for $P, R$ progressions in a number of spectra. Since each $P$, $R$ line is split into three HFS groups, the overall line intensity was determined as a sum of peak values of the groups. A couple of examples of experimental $I^{expt}_{c\to a}$ and calculated $I^{calc}_{c\to a}$ relative LIF intensity distributions are depicted in Fig.~\ref{Fig7} and Fig.~\ref{Fig7intJ50}. Fig.~\ref{Fig7} represents the intensity distribution of LIF progression for relatively high $c$ state vibrational level equal to $v_c$ = 23. The progression proceeds up to transitions to $v_a$ = 22. It can be seen that the experimental distribution is in reasonable agreement with its theoretical counterpart calculated according to Eq.~\ref{Iten}. Another LIF progression coming from the lowest $v_c$ = 0 level is depicted in Fig.~\ref{Fig7intJ50}. In spite of a very weak LIF intensity for this progression, see Fig.~\ref{Fig4J50}, the experimental intensities are in excellent agreement with calculations. Both examples unambiguously confirm the vibrational numbering of the $c$-state suggested in~\cite{Szczepkovski2017}.

\subsection{\emph{Ab initio} PECs and SO coupling functions}

\begin{figure}
\includegraphics[width=\columnwidth]{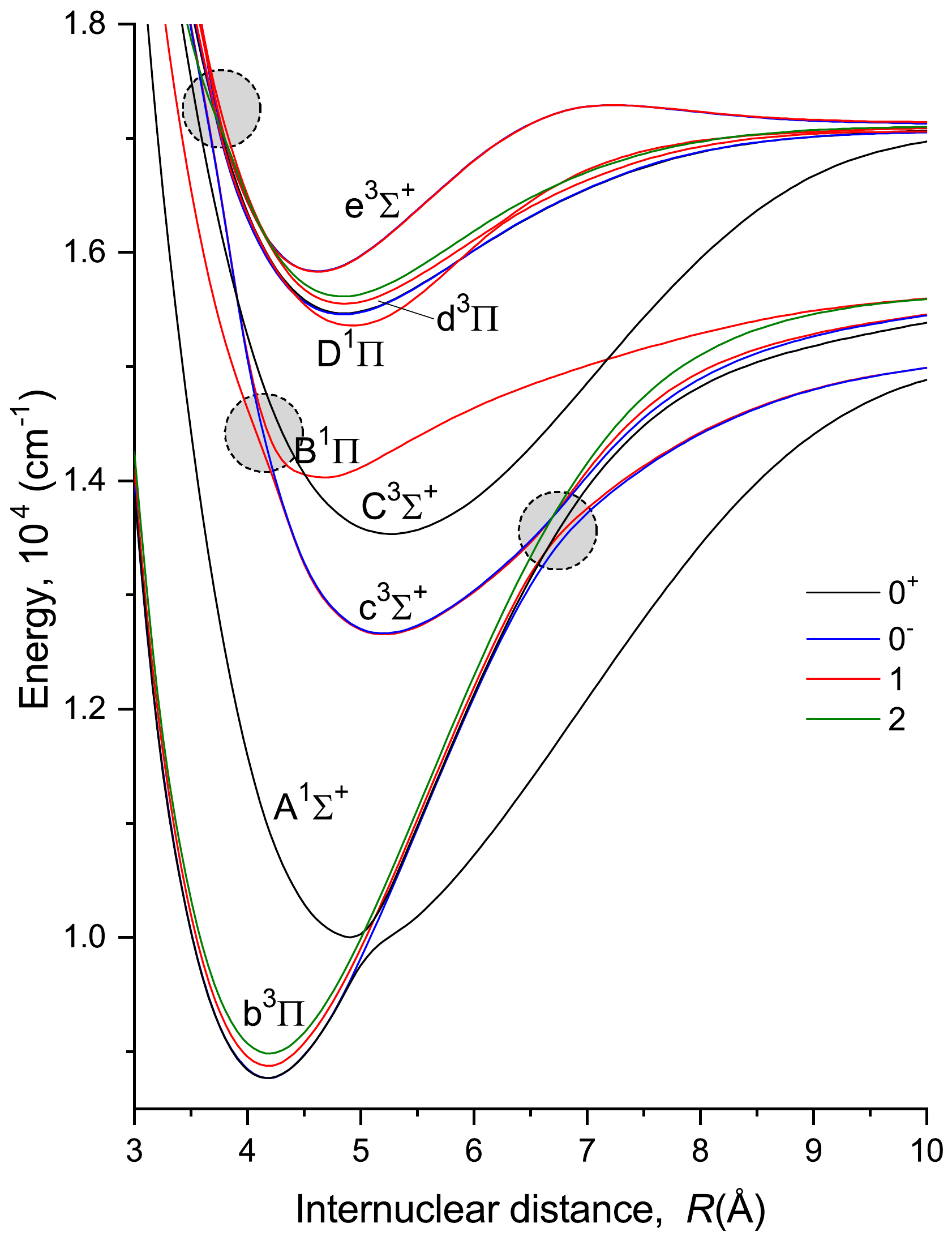}
\caption{Schema of the resulting adiabatic PECs obtained in the framework of fully relativistic FS-RCC electronic structure calculations for the KCs states converging to the second K(4$^2$S)+Cs(6$^2$P) and third K(4$^2$P)+Cs(6$^2$S) (non-relativistic) dissociation thresholds. The grey circles denote areas of the avoided crossing of the $c^3\Sigma^+$ state with $e^3\Sigma^+$, $B^1\Pi$, and $b^3\Pi$ states, respectively.}\label{FigCalcPES}
\end{figure}

In the present FS-RCC electronic structure calculations the interatomic adiabatic potentials corresponding to a pure ($c$) Hund's coupling case were obtained for all relativistic electronic states converging to the lowest three (non-relativistic) dissociation limits. The complete set of the resulting PECs is tabulated in Supplementary material and most of them is depicted in Fig.~\ref{FigCalcPES}.

In the fully relativistic framework the symmetry is lowered and electronic states are distinguished by the total angular momentum projections $\Omega$ (in our case $\Omega=0^{\pm}$, 1, and 2). This results in a number of avoided crossings between fully relativistic adiabatic states. Near the crossing points, the corresponding electronic TDMs are dramatically changed as well.
To deal with the smoothed \emph{ab initio} functions one has to transform these relativistic data to their quasi-diabatic counterparts. It can be done in the framework of the projection approach (Sect.~\ref{FSRCC}, see also~\cite{Zaitsevskii:17}). This quasi-diabatization procedure has been already outlined for the case of the $A^1\Sigma^+ \sim b^3\Pi$ complex of Rb$_2$, Cs$_2$, and RbCs molecules \cite{Krumins:20,Zaitsevskii:19}. However, the case of the $B^1\Pi \sim b^3\Pi \sim c^3\Sigma^+$ complex is much more complicated since at least three states with $\Omega=1$ are SO-coupled to each other yielding the two avoided crossing points at $R_{Bc}\approx 4.25$~\AA~ and $R_{bc}\approx 6.7$~\AA~(see Fig.~\ref{Fig2c3S}). Moreover, in the region near $R \approx 3.75$~\AA~ the $(3)1$, $(4)1$, and $(3)0^-$ states undergo one more avoided crossing with the $(5)1$ and $(4)0^-$ states belonging to the higher-lying triplet $e^3\Sigma^+$ state (Fig.~\ref{FigCalcPES}). These external avoided crossing effect (caused by both SO and radial coupling simultaneously) makes it very difficult to uncouple adiabatic curves at small internuclear separation and to obtain the quasi-diabatized PECs and corresponding effective SO coupling (SOC) matrix elements. Nevertheless, at intermediate and large distances the required smooth PECs and SOCs were obtained. Both quasi-diabatized PECs and SOC functions are available in Supplementary material.

\begin{figure}
\includegraphics[width=\columnwidth]{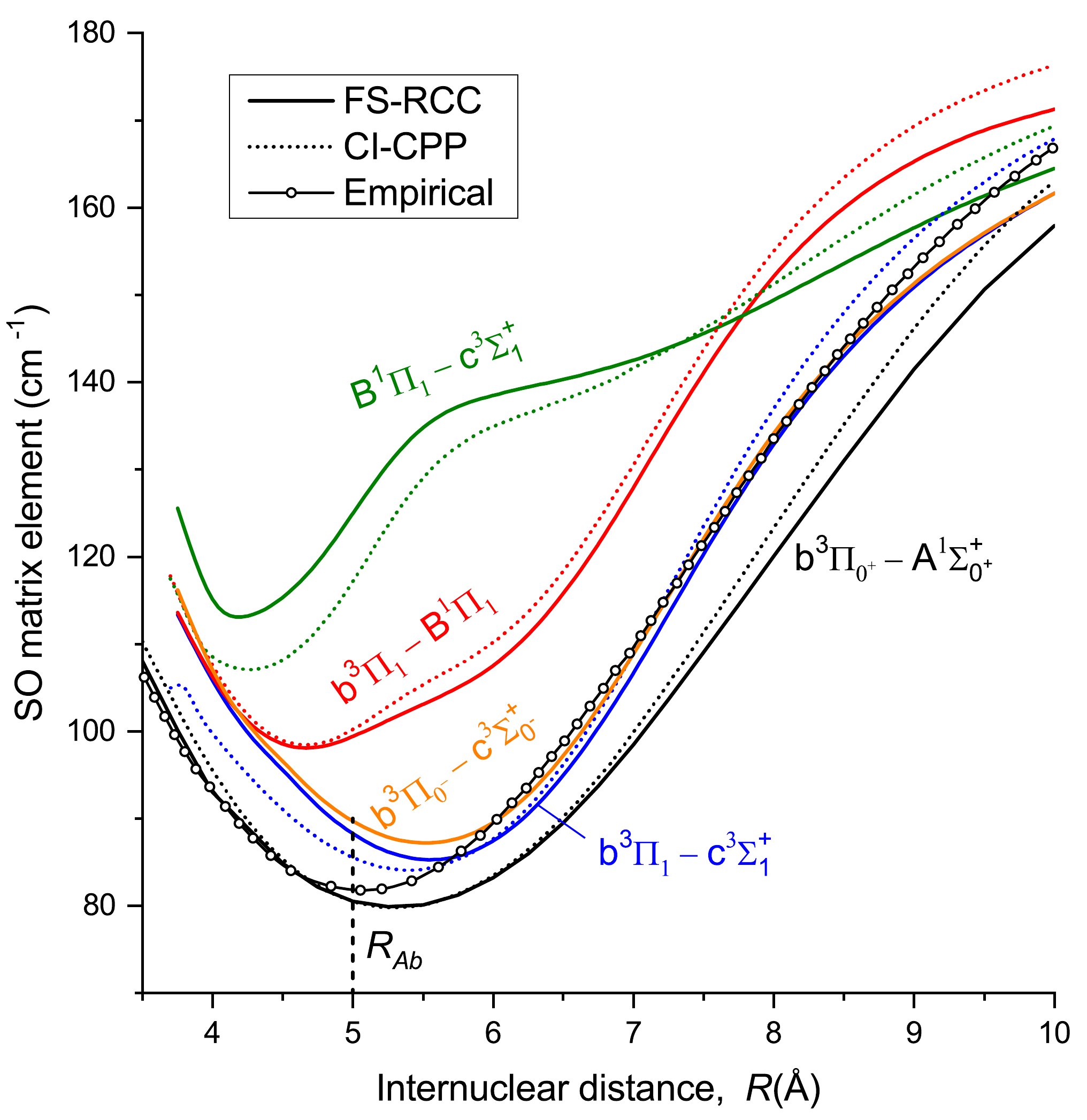}
\caption{The effective SO coupling matrix elements calculated within the present FS-RCC approach between the electronic states constituting the $B^1\Pi \sim b^3\Pi \sim c^3\Sigma^+$ and $A^1\Sigma^+ \sim b^3\Pi$ complex in KCs. The original FS-RCC $b^3\Pi_{0^-} - c^3\Sigma^+_{0^-}$ and $b^3\Pi_{0^+}-A^1\Sigma^+_{0^+}$ functions were divided by $\sqrt{2}$ to fit the definition of SOC used in previous publicationsons on their scalar-relativistic~\cite{Kim} (CI-CPP) and empirical~\cite{PRA2010} counterparts.  $R_{Ab}$ is the crossing point of the quasi-diabatizated PECs of the SO coupled $A^1\Sigma^+$ and $b^3\Pi$ states.} \label{FigCalcSO}
\end{figure}

The resulting SOC functions are depicted in Fig.~\ref{FigCalcSO}. For the $B^1\Pi \sim b^3\Pi \sim c^3\Sigma^+$ complex to the moment there are not any empirical SOC functions available constructed via the CC deperturbation analysis of experimental rovibronic term values. The empirical SOC function is available only for the $A^1\Sigma^+ \sim b^3\Pi$ complex~\cite{PRA2010,PRA2013}. The nearly perfect agreement with the previously published SOC~\cite{Kim} between the scalar-relativistic states constructed by the configuration interaction approach combined with the core polarization potential approximation (CI-CPP) should be emphasized.

The equilibrium internuclear distance $R_e$ and electronic state $T_e$ extracted from the present adiabatic and quasi-diabatic PECs are compared to their experimental and previous theoretical counterparts in Table~\ref{PECcompare}. It is clearly seen that the present \emph{ab initio} data are in overall good agreement with the experiment. In particular, the $T_e$-values derived for the the $0^-$ and 1 components of the $c^3\Sigma^+$ state predict their inverse order (see the inset in Fig.~\ref{Fig2c3S}) and the $\lambda$-splitting of the $c^3\Sigma^+$ state as $\lambda^{ab}=T_e((3)1)-T_e((3)0^-)=-11$~cm$^{-1}$. Furthermore, inserting the electronic energies of the $c^3\Sigma^+$ and $B^1\Pi$ states and the corresponding SO $c\sim B$ coupling matrix element (see Fig.~\ref{FigCalcSO}) taken at $R_e=5.2$~\AA~in Eq.(\ref{lambda}) leads to the alternative $\lambda$-estimate of 11~cm$^{-1}$. Their coincidence means that the singlet character of the triplet $c$-state (at least near the equilibrium distance) is mainly determined by its SO coupling with the nearby $B^1\Pi$ state. Using the $\lambda^{ab}$-value and assuming $S=1$ one obtains from Eq.(\ref{Qfunction1}) the value $q^{ab}_{\Omega=1}\approx +0.36$, which is only by 20\% higher than its empirical counterpart $q^{emp}_{\Omega=1}=q_0\approx +0.30$ from Table~\ref{tablq}.

It should be pointed out that the FS-RCC method excellently reproduces the shapes of the empirical potential curves in a wide range of $R$ as well. However, the systematic vertical shift of about 50-80 cm$^{-1}$ is still observed. This error can be mainly attributed to the lack of the explicit accounting for the correlation involving the $4d$ sub-shell electrons of cesium atom~\cite{Zaitsevskii:17}.

\begin{table}
\caption{Comparison of the empirical (Expt) and \emph{ab initio} (Calc) molecular constants (equilibrium distance $R_e$ and electronic energy $T_e$) available for the low-lying electronic states of the KCs molecule in the framework of both Hund's ($a$) and ($c$) coupling case representations, see indexes ($^a$) and ($^c$) in column "Source". PW - the present work.}\label{PECcompare}
\begin{center}
\begin{tabular}{cclll}
\hline\hline
 State & $\Omega$ &  Source & $R_e$~(\AA) & $T_e$~(cm$^{-1}$)\\
\hline
$c^3\Sigma^+$      & (3)$1$ & $^c$Expt[PW]                         & 5.20 & 12714 \\
                   &       & $^c$Expt\cite{Szczepkovski2017}      & 5.15 & 12720 \\
                   &       & $^c$Calc[PW]                 & 5.20  & 12653 \\
                   &       & $^c$Calc\cite{Korek2006}     & 5.10  & 12643 \\
                   & (3)$0^-$ & $^c$Calc[PW]                 & 5.20  & 12664 \\
                   &       & $^c$Calc\cite{Korek2006}     & 5.10  & 12644 \\
                   &       & $^a$Calc\cite{Habli2020}     & 5.13  & 12698 \\
                   &       & $^a$Calc\cite{Kim}           & 5.24  & 12845 \\
\hline
$B^1\Pi$           & (4)$1$   & $^c$Expt\cite{Birzniece2012,Birzniece2015}  & 4.637 & 14045 \\
                   &       & $^c$Calc[PW]                 & 4.66 & 14010 \\
                   &       & $^a$Calc\cite{Kim}           & 4.61 & 14038 \\
\hline
$b^3\Pi$           & (2)$0^+$ & $^a$Expt\cite{PRA2013}    & 4.177 & 8832 \\
                   &       & $^c$Calc[PW]              & 4.18  & 8773 \\
                   &       & $^c$Calc\cite{Korek2006}  & 4.17  & 8717 \\
                   &       & $^a$Calc\cite{Kim}        & 4.21  & 9049 \\
                   & (2)$0^-$ & $^a$Expt\cite{PRA2013}    & 4.179 & 8831 \\
                   &       & $^c$Calc[PW]              & 4.18  & 8775 \\
                   &       & $^c$Calc\cite{Korek2006}  & 4.17  & 8739 \\
                   & (2)$1$   & $^a$Expt\cite{PRA2010}    & 4.19  & 8938 \\
                   &       & $^c$Calc[PW]              & 4.18  & 8879 \\
                   &       & $^c$Calc.\cite{Korek2006}  & 4.16  & 8856 \\
                   & (1)$2$   & $^a$Expt\cite{PRA2010}    & 4.20  & 9044 \\
                   &       & $^c$Calc[PW]              & 4.19  & 8984 \\
                   &       & $^c$Calc\cite{Korek2006}  & 4.15  & 8981 \\
\hline
$A^1\Sigma^+$      & (3)$0^+$ & $^a$Expt\cite{PRA2013}       & 4.982 & 10049 \\
                   &       & $^c$Calc[PW]                 & 4.99  &  9966 \\
                   &       & $^a$Calc\cite{Kim}           & 5.02  & 10203 \\
\hline
$a^3\Sigma^+$      & (1)$0^-$ & $^c$Calc[PW]                 & 5.97 & 3797 \\
                   &       & $^c$Calc\cite{Korek2006}     & 6.02 & 3784 \\
                   & (1)$1$   & $^a$Expt\cite{Ferber2013,Ferber2009} & 6.050 & 3802 \\
                   &       & $^c$Calc[PW]                 & 5.97 & 3797 \\
                   &       & $^c$Calc\cite{Korek2006}     & 6.04 & 3785 \\
\hline
\end{tabular}
\end{center}
\end{table}

\subsection{Electronic transition moments, radiative lifetimes and vibronic branching ratios}

\begin{figure}
\includegraphics[width=\columnwidth]{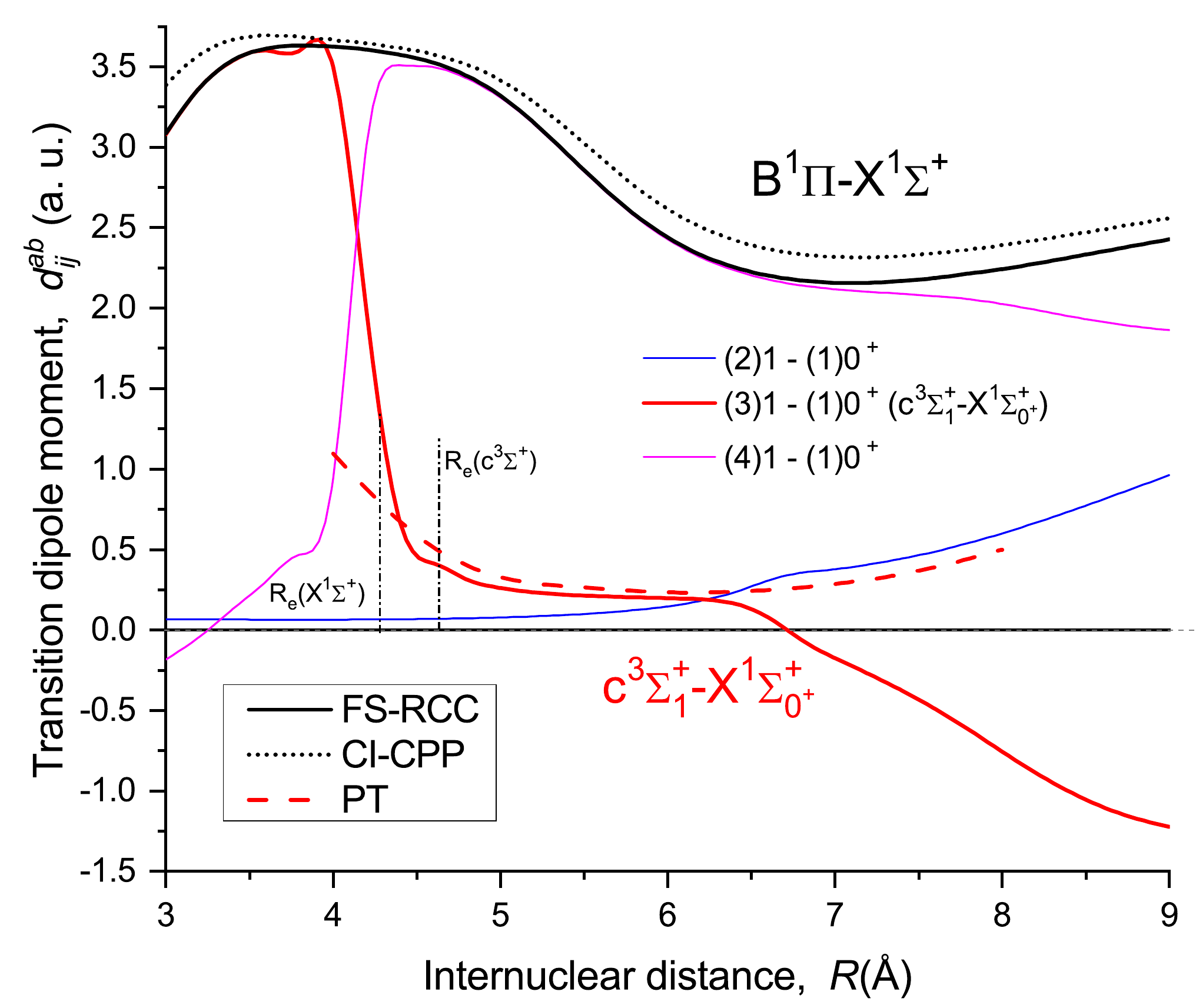}
\caption{The fully relativistic $(2,3,4)1 - (1)0^+$ TDM functions of KCs obtained in the adiabatic representation by the FS-RCC-FF method. The solid black line denotes their quasi-diabatic counterpart (FS-RCC) corresponding to the spin-allowed $B^1\Pi - X^1\Sigma^+$ transition. The red line $(3)1 - (1)0^+$ corresponds to the nominally spin-forbidden $c^3\Sigma^+_1 - X^1\Sigma^+$ transition, while the dashed red line depicts the perturbation theory (PT) counterpart evaluation by Eq.(\ref{tdmcx}). The dotted line denotes the scalar-relativistic (CI-CPP) $B-X$ TDM function borrowed from \cite{Kim}.}\label{FigCalcTDM_cXB}
\end{figure}

\begin{figure}
\includegraphics[width=\columnwidth]{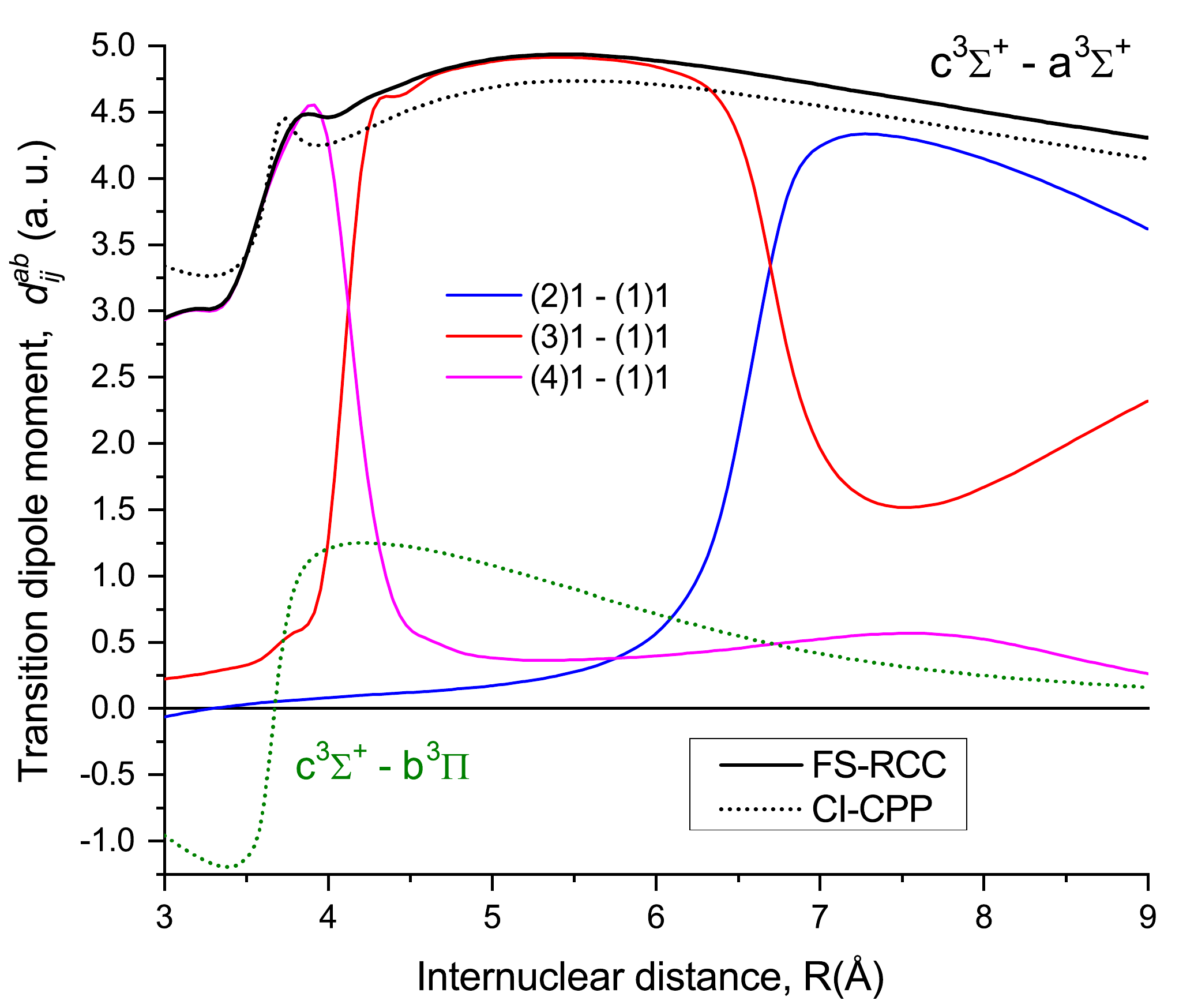}
\caption{TDM functions for the spin-allowed $c^3\Sigma^+ - a^3\Sigma^+$ and $c^3\Sigma^+ - b^3\Pi$ transitions in the KCs molecule. The fully relativistic $(2,3,4)1 - (1)1$ curves are obtained in the adiabatic representation and then transformed to their quasi-diabatic counterpart $c^3\Sigma^+ - a^3\Sigma^+$ (solid black line). Dotted line denotes the relevant scalar-relativistic CI-CPP curve borrowed from Ref.~\cite{Kim}.}\label{FigCalcTDM_ca}
\end{figure}

In order to simulate radiative properties of the relativistic states under study the relevant electronic TDM functions have been also calculated within the same FS-RCC method combined with a finite-difference approach (see Sec.~\ref{FSRCC}). In the present work we have focused on the radiative properties of the $c^3\Sigma^+$ state of KCs. We start from the spin-forbidden transition from/to the ground $X^1\Sigma^+$ state. This transition was used both in the present and previous work~\cite{Szczepkovski2017} to excite the triplet $c^3\Sigma^{+}_{1^\pm}$ state directly from the ground singlet state. The resulting $d^{ab}_{cX}(R)$ function corresponding to the fully relativistic $(3)1-(1)0^+$ transition is depicted in Fig.~\ref{FigCalcTDM_cXB}. It can be seen that the $(3)1-(1)0^+$ adiabatic transition moment $d^{ab}_{cX}$ is about 0.2-0.7~a.u. in the range of $R\in[4.2,6.5]$~\AA~ making the nominally spin-forbidden $c \leftarrow X$ transition to be allowed, obviously due to the admixture of the nearby singlet $B^1\Pi$ state:
\begin{eqnarray}\label{tdmcx}
d^{PT}_{cX}\approx \frac{d^{ab}_{BX}\xi^{so}_{Bc}}{U_c-U_B}
\end{eqnarray}

To obtain the required smooth TDM function for the spin-allowed $c^3\Sigma^+ - a^3\Sigma^+$ transition one has to transform the relevant adiabatic TDMs to their quasi-diabatic counterpart. We have done that imposing the obvious conditions for formally spin-forbidden transition $\langle B^1\Pi_1|\hat{d}|a^3\Sigma^+_1\rangle = 0$ and the longitudinal transition $\langle b^3\Pi_1|\hat{d}_{||}|a^3\Sigma^+_1\rangle = 0$. In doing so, one arrives at a simple expression:
\begin{eqnarray}\label{diab-tdm-ca}
|\langle c^3\Sigma^+_1|\hat{d}|a^3\Sigma^+_1 \rangle|^2 = \sum_{i=2,3,4}|\langle (i)1|\hat{d}|(1)1\rangle|^2,
\end{eqnarray}
which is not exact since it is based on the assumption that the states to be diabatized are pure spin states. The related TDM functions are depicted in Fig.~\ref{FigCalcTDM_ca}. The resulting $c^3\Sigma^+ - a^3\Sigma^+$ TDM is in excellent agreement with the previous scalar-relativistic CI-CPP calculation~\cite{Kim}.

The high accuracy of the present fully relativistic TDM functions is confirmed by the fact that at dissociation limit they converge to the value 3.22 a.u. The corresponding experimental atomic value for the non-relativistic $6^2S - 6^2P$ transition in Cs is equal to 3.1822(18) a.u.~\cite{Sansonetti:05}. Thus, the deviation of the FS-RCC-FF value from the experimental one does not exceed 1.5\%.

\begin{figure}
\includegraphics[width=\columnwidth]{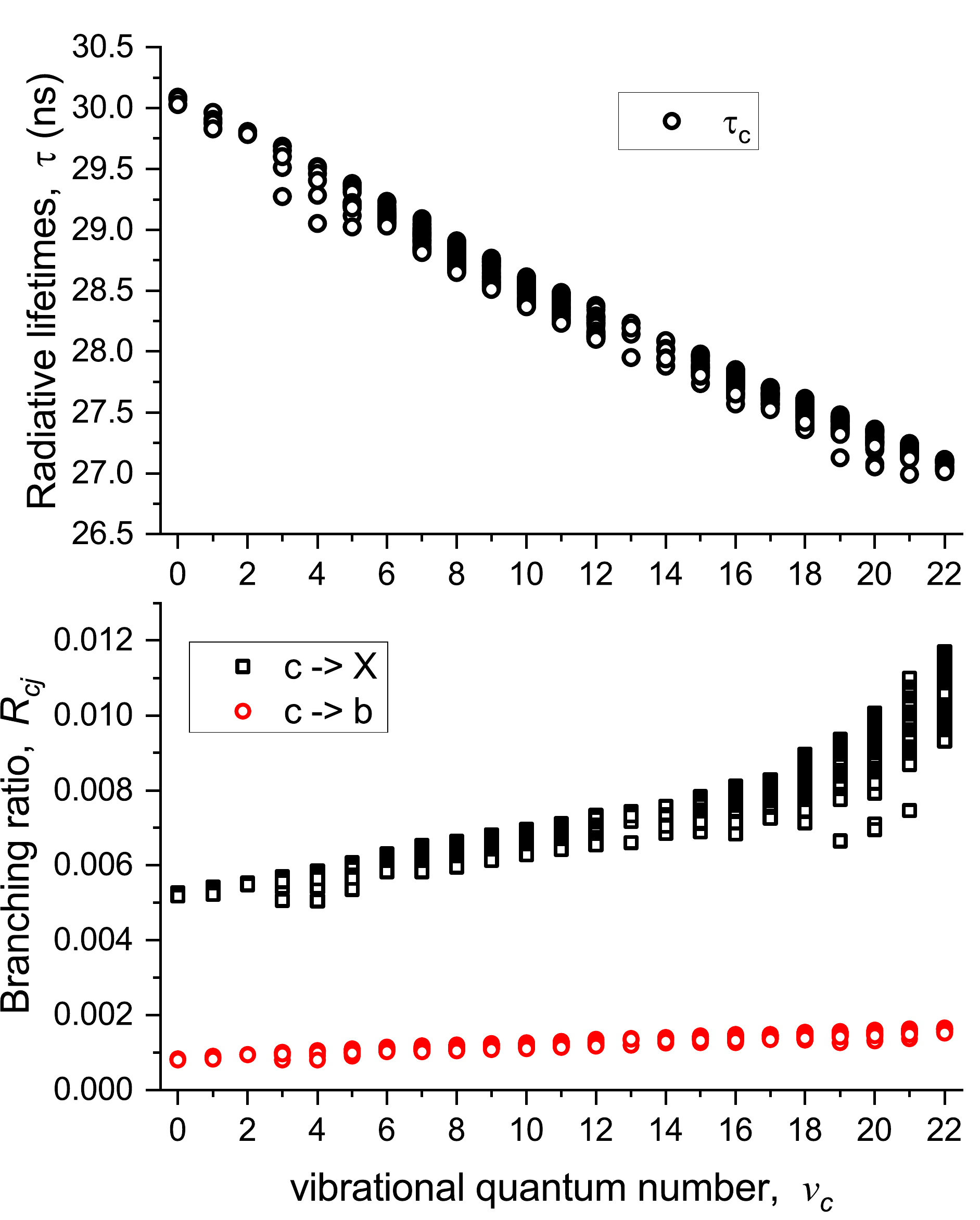}
\caption{Radiative lifetimes, $\tau_c$, and vibronic branching ratios, $R_{cX}$ and $R_{cb}$, predicted for the experimentally observed rovibronic levels of the KCs $c^3\Sigma^+(v_c\in[0,22])$ state.} \label{FigCalctau}
\end{figure}

The derived FS-RCC-FF transition dipole moment functions were substituted into Eqs.(~\ref{tausum}) in order to estimate radiative lifetimes $\tau_c$, and vibronic branching ratios $R_{ca}$ and $R_{cb}$, for the lower-lying rovibronic levels of the $c^3\Sigma^+$ state experimentally observed in the present work, as well as in~\cite{Szczepkovski2017} and applied here for the refined IPA potential construction. The resulting values are depicted in Fig.~\ref{FigCalctau}. As expected, the dominant contribution (about 99\%) into the simulated $\tau_c$-values is coming from the spin-allowed $c\to a$ transition. The fraction part of the rest $c\to X$ and $c\to b$ transitions does not exceed 1\% and 0.1\% of the main transition, respectively. The calculated lifetimes smoothly decrease from 30 to 27 ns as $v_c$-values increases from 0 to 22, whereas the branching ratio $R_{cX}$ increase from 0.005 until 0.01 in the same $v_c$-range.

\section{Conclusions}

We performed a high-resolution Fourier-transform spectroscopic study of the $c^3\Sigma^+(\Omega=1^{\pm}, 0^{-})$ state of the KCs molecule by recording and analyzing the spin-allowed $c^3\Sigma^{+}_{1^\pm} \rightarrow a^3\Sigma^{+}_{0^-,1^\pm}$ fluorescence spectra excited by formally spin-forbidden laser-induced $c^3\Sigma^{+}_{1^\pm} \leftarrow X^1\Sigma^+_{0^+}$ transitions.

A list of rovibronic energies $E_c$ of the obtained experimental term values is presented in Table I of Supplementary materials; the accuracy is estimated within 0.010 and 0.015 cm$^{-1}$. The direct potential fit was performed on present experimental term values in the interval $v^{\prime}_c\in [0,22]$, $J^{\prime}_c\in [11,141]$, that included 201 $e$-levels and 54 $f$-levels supplemented by 469 $e$-parity term values from Ref.~\cite{Szczepkovski2017}. The resulting refined interatomic IPA potential has reproduced the present empirical term values within experimental accuracy, as can be seen from the Fig.~\ref{Fig6resAS5} and also from Table II of Supplementary material. This provides a possibility to predict the laser frequency on demand for excitation of a particular rovibronic level of the $c$-state. Mass-invariant properties of the derived IPA potential are confirmed by a good agreement of the predicted and experimental rovibronic term value for the $^{41}$KCs isotopologue. Observation of rovibronic level with $v_c$ = 0 allowed to more precisely evaluate the $T_e$-value of the $c$-state, that appeared to be by about 6 cm$^{-1}$ lower than the respective value predicted by~\cite{Szczepkovski2017}. The present empirical $R_e$ value is also expected to be more accurate; as seen from Table II, it coinsides with \emph{ab initio} calculations. For higher $v_c$ values, the potential from~\cite{Szczepkovski2017} can be useful.

Detection of $e$- and $f$-components of the $c^3\Sigma^{+}_{1^\pm}$ state made it possible to explicitly account for $\Omega$-doubling between the $\Omega=1^+$ and $\Omega=1^-$-components; the experimental $\Omega$-doubling constant value is about 1$\times$10$^{-4}$ cm$^{-1}$. A considerable amount of presentlt determined high accuracy term values for vibrational levels $v^{\prime}_c>22$ contains valuable information about $c-B$ and $c-b$ perturbations.

Interatomic adiabatic and quasi-diabatic potentials of the $A^1\Sigma^+$, $B^1\Pi$, $b^3\Pi$ and $c^3\Sigma^+$ states that can be referred to as the spin-orbit coupled $A-B-b-c$ complex of the states converging to the second non-relativistic dissociation limit have been calculated (along with the related effective SO couplings electronic matrix elements) in the framework of the Fock space relativistic coupled-channel approach. Though the required rigorous couple-channel deperturbation analysis of both $e$- and $f$-components of the $A-B-b-c$ complex up to a common dissociation limit looks right now to be too complicated task, we believe that the present experimental and theoretical studies will greatly facilitate to perform this challenging treatment in near future.

The comparison of relative intensity distributions obtained from the recorded $c\to a$ LIF spectra with the calculated $c\to a$ rovibronic transition probabilities  demonstrated good agreement, thus unambiguously confirming the $c$-state absolute vibrational numbering and supplying an additional proof of the correctness of energy-based properties.

The electronic transition dipole moments between the lowest excited and ground states of KCs have been \emph{ab initio} calculated using the finite-field approach combined with the FS-RCC method. The singlet feature of the nominally triplet $c^3\Sigma^+$ state is mainly borrowed from the nearby $B^1\Pi$ state. The resulting spin-allowed $c-a$, $c-b$ and spin-forbidden $c-X$ transition moments were used to estimate the radiative properties of the $c^3\Sigma^+$ rovibronic levels $v_c\leq 20$. The possible systematic uncertainty in the predicted radiative lifetimes of the $c$-state and branching ratios of $c\to a$, $c\to b$ and $c\to X$ emission transitions should, as a rule, not exceed 3-4\%.

\section*{Acknowledgements}
We are grateful to V. V. Meshkov for providing the FORTRAN subroutine generating the highly accurate empirical CPE interatomic potentials for the ground singlet and triplet states of KCs.
Moscow team is grateful for the support by the Russian government budget (section 0110), projects No.121031300173-2 and 121031300176-3. Riga team acknowledges the support from the Latvian Council of Science, project No. lzp-2018/1-0020: "Determination of structural and dynamic properties of alkali diatomic molecules for quantum technology applications" and from the University of Latvia Base Funding No A5-AZ27.

\end{document}